\UseRawInputEncoding
\documentclass[twocolumn,showpacs,preprintnumbers,amsmath,amssymb]{revtex4}
\usepackage{graphicx}% Include figure files
\usepackage{bm}% bold math

\usepackage{epsf}
\usepackage{epstopdf}
\begin{document}
%\draft
\title{A new mechanism of structural transition in 2D Hertzian spheres in the presence of random pinning }%:
%First-Order versus Continuous Transition }

\author{E. N. Tsiok, Yu. D. Fomin, E. A. Gaiduk, and V. N. Ryzhov}
\affiliation{ Institute of High Pressure Physics RAS, 108840
Kaluzhskoe shosse, 14, Troitsk, Moscow, Russia}

\date{\today}

\begin{abstract}
Using molecular dynamics simulation we have investigated the
influence of random pinning on the phase diagram and melting
scenarios of a two-dimensional (2D) system with the Hertz
potential for $\alpha=5/2$. For the first time it has been shown
that random pinning can cardinally change the mechanism of
first-order transition between the different crystalline phases
(triangular and square) by virtue of generating hexatic and
tetratic phases: a triangular crystal to hexatic transition is of
the continuous Berezinskii-Kosterlitz-Thouless (BKT) type, a
hexatic to tetratic transition is of the first-order, and finally,
a continuous BKT type transition from tetratic to the square
crystal.
\end{abstract}


\maketitle

\section{Introduction}

Studying the self-organization of 2D systems, especially
soft/deformable colloidal mesoparticle systems such as dendrimers,
star polymers, and block-copolymer micelles \cite{likos,c1,c2}, is
of great interest for both fundamental science and technological
applications. Of particular interest for optical applications are
the structures of crystalline phases and the relation of these
structures with the form of interparticle potential. The most
popular potentials used to describe deformable nanocolloids are
nontrivial phenomenological interactions, some of which even lead
to a complete overlap among the components and demonstrate very
rich phase behavior \cite{likos,c4,c13,c14,c15}. It seems that the
simplest form of the family of these potentials is the Hertz
potential \cite{c4}. Recently, the behavior of 2D Hertzian spheres
has been studied in a number of articles
\cite{softmat,molphys,miller,hertzmelt,hertzqc}.

In the present article we are going to discuss the influence of
disorder on the phase diagram of 2D Hertzian spheres. As far back
as in the 70s it was established that the melting of 2D systems
could in principle be different from the melting of
three-dimensional (3D) crystals. If in the 3D case melting is
always a first-order phase transition then 2D systems can melt
according to several different scenarios (see \cite{pu2017,pu2020}
and the references in these works). Today there are known at least
three different melting scenarios of 2D systems: 1) melting via a
first-order phase transition \cite{chui,ryzhovJETP}; 2) a melting
scenario according to the
Berezinskii-Kosterlitz-Thouless-Halperin-Nelson-Young theory
(BKTHNY) \cite{ber1,kostth,halpnel1,halpnel2,young}. In this
scenario melting takes place via two continuous transitions of the
BKT type. As a result of the first transition, the long-range
orientational order is destroyed in the crystal and transforms
into quasi-long-range (power decay of the orientational order
correlation functions) and the translational order from
quasi-long-range becomes short-range. The obtained phase is called
hexatic. The second continuous BKT phase transition leads to a
disappearance of the quasi-long-range orientational order, as a
result of which the system changes to isotropic liquid with
short-range orientational and translational orders. Finally, the
third melting scenario of 2D crystals is as follows: melting also
occurs in two stages but transition from crystal to hexatic is
continuous of the BKT type, and from hexatic to liquid is of the
first order \cite{foh1,foh2,foh3,foh4,hs}. We will call these
scenarios as the first, the second and the third.

It is significant that in all classical works on studying 2D
melting only one crystalline structure was considered, i.e., a
triangular crystal that is a close-packed structure in two
dimensions. At the same time, recent experimental works have shown
the possibility of existence of other 2D and quasi-2D crystalline
structures. A more wide-known example is graphene that is a sheet
of graphite, i.e., a 2D layer with a honeycomb structure
\cite{graphene}. Later on, other 2D and quasi-2D structures were
also found, for instance, square ice in water confined in a slit
pore \cite{geim}, a square crystal of iron atoms in the defects of
graphene \cite{iron}, complex crystalline structures in a thin
colloidal film \cite{dobnikar} and in a system of vortices in
superconductors \cite{sc1,sc2}. However up to now obtaining
non-triangular 2D crystals is rather an exception than a rule and
the overwhelming majority of 2D systems crystallize exactly into a
triangular lattice.

At the same time, non-triangular 2D crystalline lattices have been
found in a large number of works on computer simulation of 2D
systems, for instance, in a system with two scale repulsive
shoulder potentials \cite{Franzese, Rice1, Rice2, PhysicaA,
sc3,sc4} and in water \cite{Franzese1,
Franzese2,barb1,barb2,Rice3}. In the works
\cite{dfrt1,dfrt2,dfrt3,jain,marcotte} a square crystal was found,
in \cite{jain,marcotte} a honeycomb structure was observed, the
Kagome lattice was discovered in \cite{pineros, Rice3}. In a
number of publications formation of quasi-crystalline phases in 2D
systems was reported \cite{qc1,qc2,qc3,we5}.

Experimental investigation of 2D crystal melting is complicated by
the presence of the so-called "pinning" of particles. It means
that because of the effects of interaction with the underlayer
some particles turn out to be pinned to certain fixed places. It
is clear that the presence of pinning and the concentration of
pinned particles may considerably affect a system's behavior.

One should discriminate between the two types of pinning. In the
case of quenched disorder, a random fraction of particles could be
pinned either to random positions in the system or on lattice
sites of the underlying crystal phase. With regard to the pinned
particles at random sites, it was shown theoretically that
quenched disorder influenced crystalline order, but almost did not
affect orientational order. So the BKTHNY melting scenario
persisted, and the solid phase was destroyed entirely for high
pinning fractions (see \cite{nels1, nels2, cha, grun, soft-pin,
dfrt6}). Experiments and simulations of 2D melting of
superparamagnetic colloidal particles with quenched disorder
confirmed the increased stability range of the hexatic phase (see
\cite{keim2,keim3,zahn}).

However, in the paper \cite{Dijkstra1} the authors study the
melting of a 2D system of hard disks with quenched disorder, which
results from pinning random particles on a crystalline lattice.
This kind of pinning stabilizes the solid phase and can destroy
the hexatic phase. We are not aware of the real experiments with
this kind of disorder.

Finally, in \cite{dfrt5} it was demonstrated, that random pinning
could qualitatively change the first-order melting scenario - it
can generate the hexatic phase and transform the first-order
transition into the third type melting scenario.

In our article we investigate a system with particles pinned at
random sites, including interstitial lattice sites. It should be
noted that previously the influence of pinning on melting of only
the triangular crystal was studied. As far as we know, the
influence of pinning on melting of other 2D crystals has not been
explored yet.

One of the systems, in which in the 2D case a complex phase
diagram with a large number of different phases is observed, is
the Hertz model. It is a system of particles that interact through
the potential

\begin{equation}\label{pot}
   U(r)=\varepsilon \left ( 1- r/ \sigma \right)^{\alpha}H(1-r),
\end{equation}
where $H(r)$ is the Heaviside step function and parameters
$\varepsilon$ and $\sigma$ set the energy and length scales. In
the case of $\alpha=5/2$ the Hertz potential corresponds to the
energy of deformation of two elastic spheres \cite{landau_el}.

The phase diagram of Hertzian spheres with $\alpha=5/2$ has many
different ordered phases in both three dimensions
\cite{hertz3d,rosbreak} and two dimensions \cite{molphys}.
Besides, in Hertzian spheres a number of anomalous properties of
liquid are observed (see \cite{rosbreak} for the 3D case and
\cite{molphys} for the 2D one).

The phase diagram of 2D Hertzian spheres with $\alpha=5/2$ was
discussed in several publications \cite{miller,hertzmelt,molphys}.
The most complete calculation of this phase diagram is given in
the work \cite{molphys}, which shows that in this system several
stability regions of the triangular crystal are observed, several
of the square one as well as a number of other phases including
the dodecagonal quasicrystal. Besides, in that work the melting
scenarios of the triangular and square crystals with low density
were determined and it was shown that in this system all three
currently known melting scenarios took place. During changing from
low to high densities, the melting lines of both triangular and
square lattices pass through a maximum, after which the melting
temperature begins falling with an increase in density. Moreover,
in the region of the reentrant melting curve of the triangular
phase there are two tricritical points, in which a change in the
melting scenarios takes place from the third to BKTHNY at the
maximum and from BKTHNY to the third at lower temperatures. The
tricritical point on the melting curve of the square crystal is
located at the maximum, in which a change in the transition
scenario from the third to a first-order transition takes place.

\begin{figure}
\includegraphics[width=8cm]{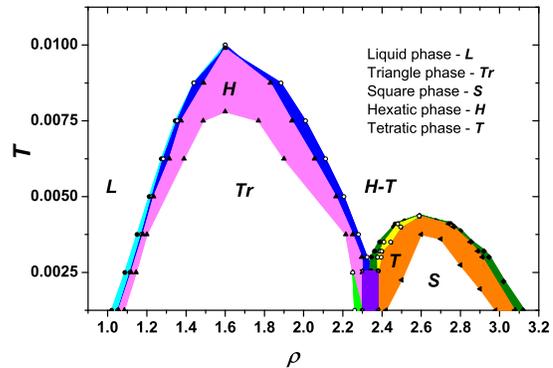}%

\caption{\label{pd-pin} The full phase diagram of a pure system
and of a system with pinning in the area of existence of the
triangular and square crystals. 1) cyan - is liquid-hexatic phase
coexistence; 2) blue - is the hexatic phase without pinning; 3)
magenta - is the hexatic phase with 0.1\% pinning; 4) green - is
the extension of the hexatic phase for 0.2\% pinning; 5) olive -
is liquid-tetratic phase coexistence; 6) yellow - is the tetratic
phase without pinning; 7) orange - is the tetratic phase with
0.2\% pinning; 8) violet - is hexatic-tetratic coexistence. See
explanations in the text.}
\end{figure}

In the present paper we examine the influence of random pinning on
the melting scenario of the triangular and square crystals of 2D
Hertzian spheres with $\alpha=5/2$ and on the transition between
these two crystalline phases in the region of reentrant melting at
low temperatures. The influence of random pinning on the
structural transition has never been considered up to now. As
mentioned above, the previous works only considered the influence
of random pinning on melting of the triangular crystal where
melting occurs according to the first and the third scenario
\cite{dfrt5,soft-pin}. Therefore studying the behavior of 2D
Hertzian spheres at different concentrations of random pinning
allows us to solve three problems at once: 1) considering the
influence of random pinning on melting of the triangular crystal
when the different melting scenarios take place and 2)
investigating the influence of random pinning on melting of the
square crystal in the two different melting scenarios. At lat, we also
demonstrate that random pinning drastically changes the scenario
of the structural transition between the triangular and square
crystalline phases. In Fig.~\ref{pd-pin} we summarize the results
which will be discussed in detail below.

\section{System and methods}

In the present paper using a molecular dynamics method within the
framework of the software package LAMMPS \cite{lammps} we
simulated a 2D system of Hertzian spheres in the region of low
densities, in which the triangular and square phases were observed
(see Fig.~\ref{pd-pin}). When investigating the triangular phase,
{N}=20000 particles were simulated in a rectangular box, whereas
to study the square phase - {N}=22500 particles in a square box
were used. In all cases, periodic boundary conditions were
applied. The samples of liquid were obtained by melting respective
lattices.

The systems were simulated using 60 million steps with time
interval $dt=0.0001$ in NVT- and NVE-ensembles (NVT - for
equilibration and NVE - for production), the first 30 million
steps were used for equilibration. The concentration of pinning
particles at random positions varied from 0.1\% (the triangular
lattice) to 0.2\% (the square lattice). The procedure of
introduction of pinning particles is described in detail in
\cite{dfrt5,soft-pin,dfrt6}. We considered not fewer than 10
different systems with various initial random positions of fixed
particles with further averaging by these replicas. In the course
of simulation, the system pressure was calculated as a function of
density and temperature. The transition regions between different
phases were determined based on the peculiarities on the equations
of state (isotherms) - the Mayer-Wood loop (an analogy of the van
der Waals loop for the 3D case), while the structure of these
phases was obtained from the radial distribution functions. We
also calculated orientational and translational order parameters
and their correlation functions to determine the stability limits
of the triangular and hexatic phases as well as the square and
tetratic (an analogy of hexatic for the square crystal) phases.

The translational order parameter is calculated in the standard
way \cite{halpnel1,halpnel2,dfrt5,dfrt6}:
\begin{equation}
\psi_{t}=\frac{1}{N}\left<\left<\left|\sum_{j} e^{i{\bf G
r}_j}\right|\right>\right>_{rp}\label{psit},
\end{equation}

The local orientational parameter is given in the following way
\cite{halpnel1,halpnel2,dfrt5,dfrt6}:
\begin{equation}
\Psi_{mj}=\frac{1}{N_{j}}\sum_{k=1}^{N_{j}} e^{mi
\theta_{jk}}\label{psi6loc},
\end{equation}
where ${N_{j}}$ - is the number of the nearest neighbors of
particle {j} that is determined from the Voronoi construction,
$\theta_{jk}$ is the angle of the bond between particles $j$ and
$k$ relative to an arbitrary reference axis, $\bf G$ is a primary
reciprocal lattice vector, and index $rp$ points to averaging by
10 replicas with different initial random positions of the fixed
particles. The global orientational order parameter is obtained by
means of averaging over all particles
\begin{equation}
\psi_m=\frac{1}{N}\left<\left<\left|\sum_{j}^{N} \Psi_{mj}\right|\right>\right>_{rp}\label{psi6},
\end{equation}
where $m=6$ for the triangular lattice and $m=4$ for the square
one.

The orientational correlation function (OCF) is defined as
\begin{equation}
G_m(r)=\left<\frac{\left<\Psi_m({\bf r})\Psi_m^*({\bf 0})\right>}{g(r)}\right>_{rp},
\label{g6}
\end{equation}
where $g(r)=<\delta({\bf r}_i)\delta({\bf r}_j)>$ is the pair
distribution function. In the hexatic and tetratic phases the long
range behavior of $G_m(r)$ has the form $G_m(r)\propto
r^{-\eta_m}$ with $\eta_m \leq \frac{1}{4}$ \cite{halpnel1,
halpnel2}.

The translational correlation function (TCF) is calculated as
\begin{equation}
G _t(r)=\left<\frac{<\exp(i{\bf G}({\bf r}_i-{\bf r}_j))>}{g(r)}\right>_{rp},
\label{GT}
\end{equation}
where $r=|{\bf r}_i-{\bf r}_j|$. In the solid phase the long range
behavior of $G_t(r)$ has the form $G_t(r)\propto r^{-\eta_T}$ with
$\eta_T \leq \frac{1}{3}$ \cite{halpnel1, halpnel2}. The stability
limits of the square crystal were determined in the same way. In
the hexatic/tetratic phase and isotropic liquid $G_t(r)$ decays
exponentially.

The presence of random pinning decreases the range of stability of
a crystalline phase and consequently relatively expands the area
of existence of the hexatic/tetratic phase, which allows
investigating its dynamic properties by means of calculating the
diffusion coefficient \cite{dfrt6}. To do this Einstein's method
was used, i.e., mean-square displacement $<r^2(t)>$ was
calculated, that is proportional to time at large times:
$<r^2(t)>=4Dt$, where $D$ is a diffusion coefficient.

%\section{Results and discussion}

\section{Melting of the triangular crystal in the presence of random pinning}

Let us consider Hertzian spheres with random pinning $0.1 \%$ in
the melting area of the triangular crystal (see
Fig.~\ref{pd-pin}). Recall that random pinning, as a rule, does
not practically affect the stability limit of hexatic but
significantly decreases the stability area of a crystal. As a
result, compared with a system without random pinning the area of
existence of the hexatic phase relatively increases \cite{molphys,
soft-pin, nels1, nels2, cha, keim3, dfrt5}.

\begin{figure}
\includegraphics[width=8cm]{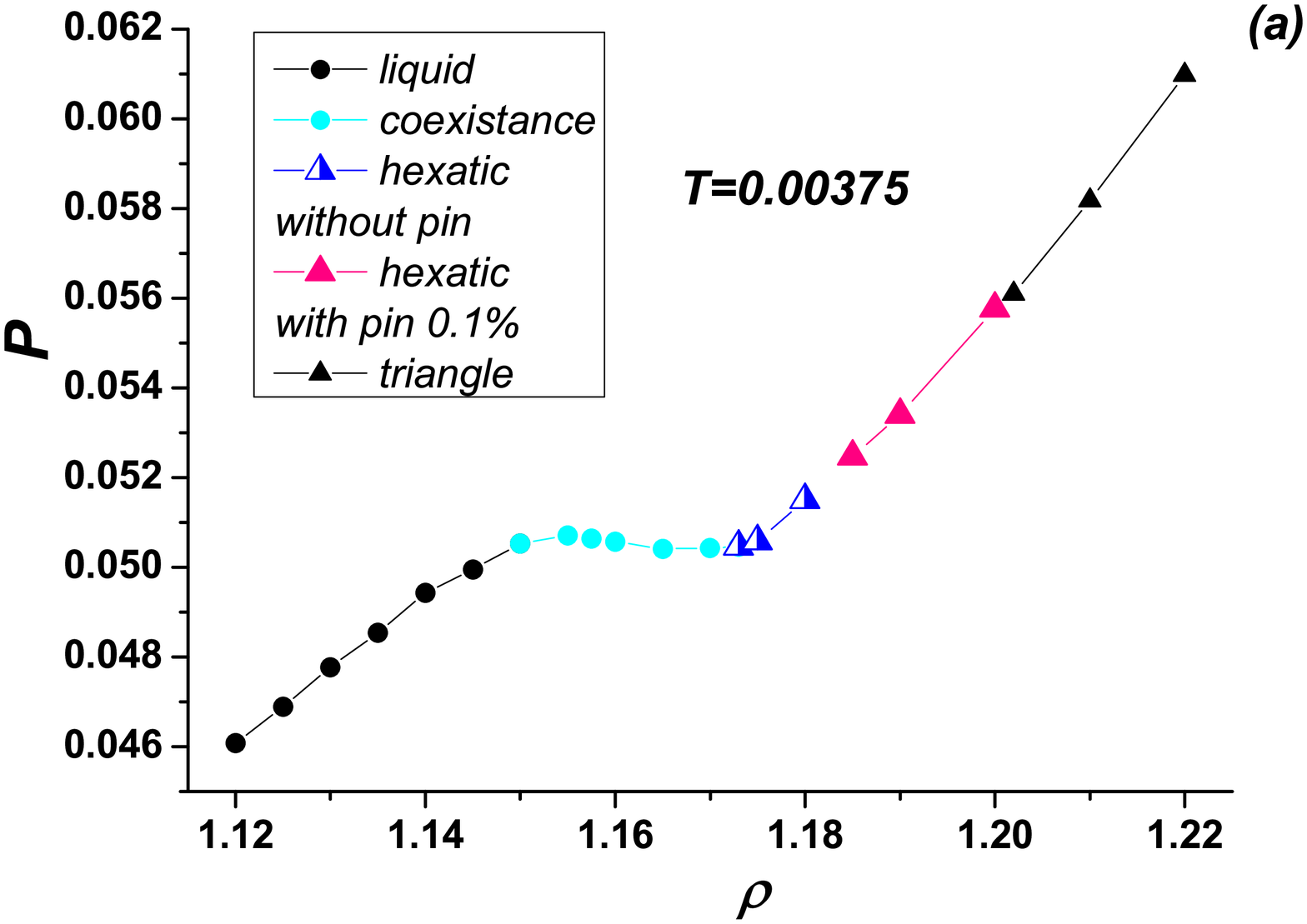}%

\includegraphics[width=8cm]{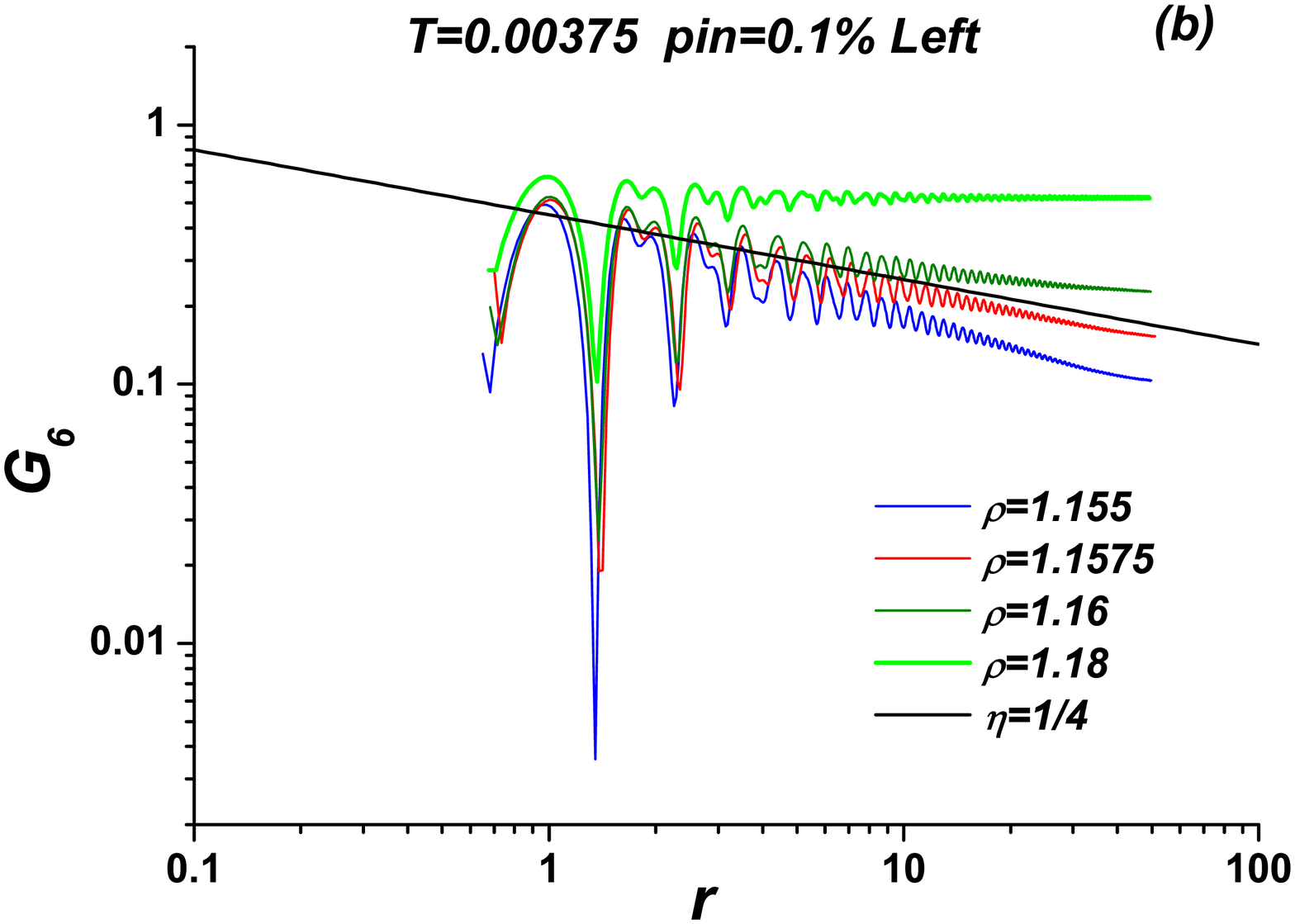}%

\includegraphics[width=8cm]{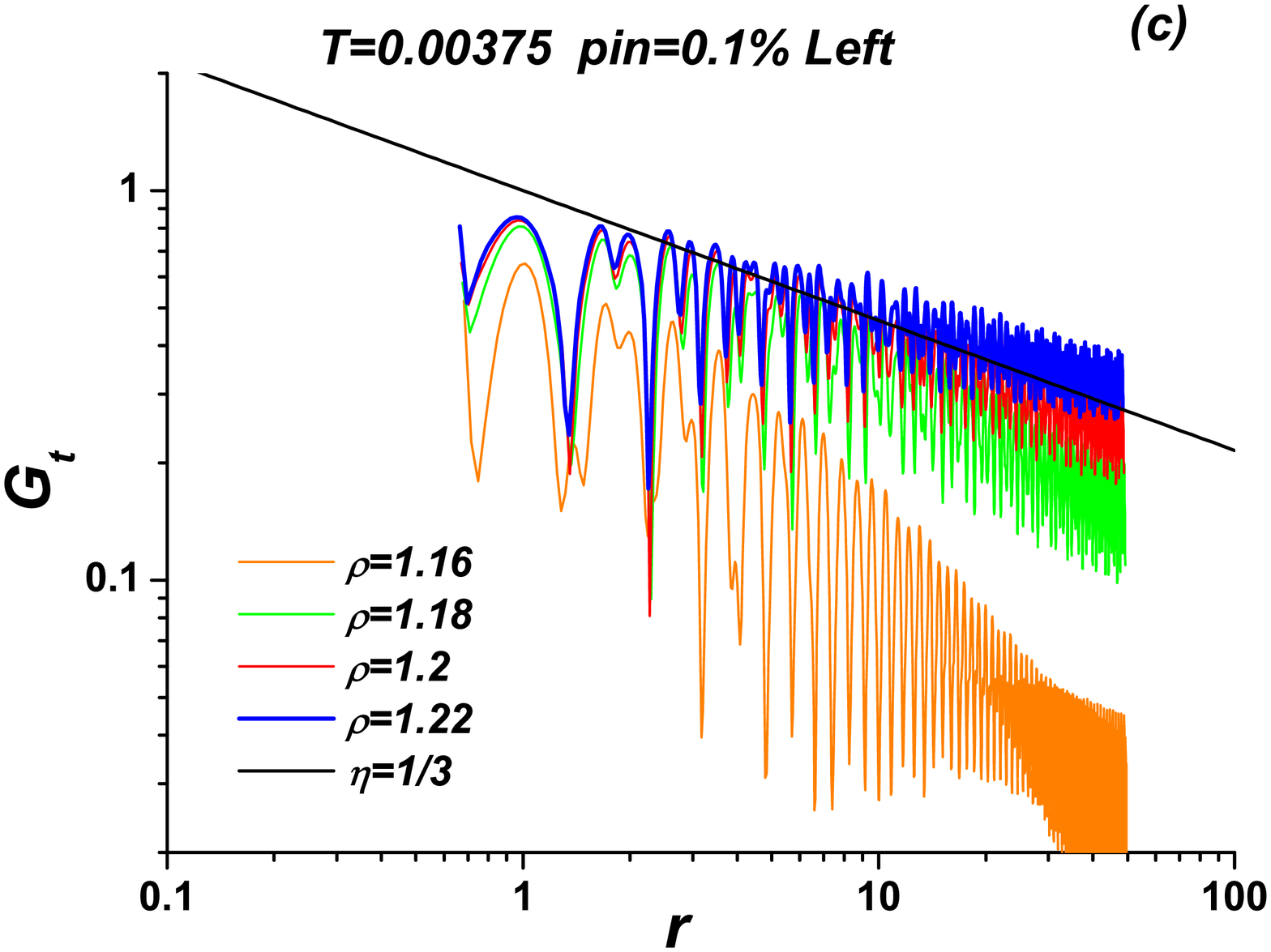}%

\caption{\label{tr-left} (a) The equation of state of Hertzian
spheres at $T=0.00375$ with concentration of random pinning $0.1
\%$ at the left branch of the triangular crystal melting line. (b)
The behavior of OCF $G_6$ of the same system. (c) The behavior of
TCF $G_t$ of the same system.}
\end{figure}

Fig.~\ref{tr-left} (a) presents the system's equation of state on
isotherm $T=0.00375$ at the crossing of the left branch of the
triangular crystal melting curve with the Mayer-Wood loop
characteristic of first-order transition. On this isotherm we
marked the points of stability loss of the crystal and hexatic
obtained from analysis of the orientational and translational
order parameter correlation functions shown in Fig.~\ref{tr-left}
(b) and (c). From the obtained results it can be concluded that
the crystal to hexatic transition is a continuous one of the BKT
type, while the hexatic to isotropic liquid transition is a
first-order transition. Thus, on the left branch the melting
scenario remained the same as in the system without pinning (the
third scenario) \cite{molphys} but the stability limit of the
crystal shifted to higher densities, which led to expansion of the
hexatic phase existence area.

\begin{figure}
\includegraphics[width=8cm]{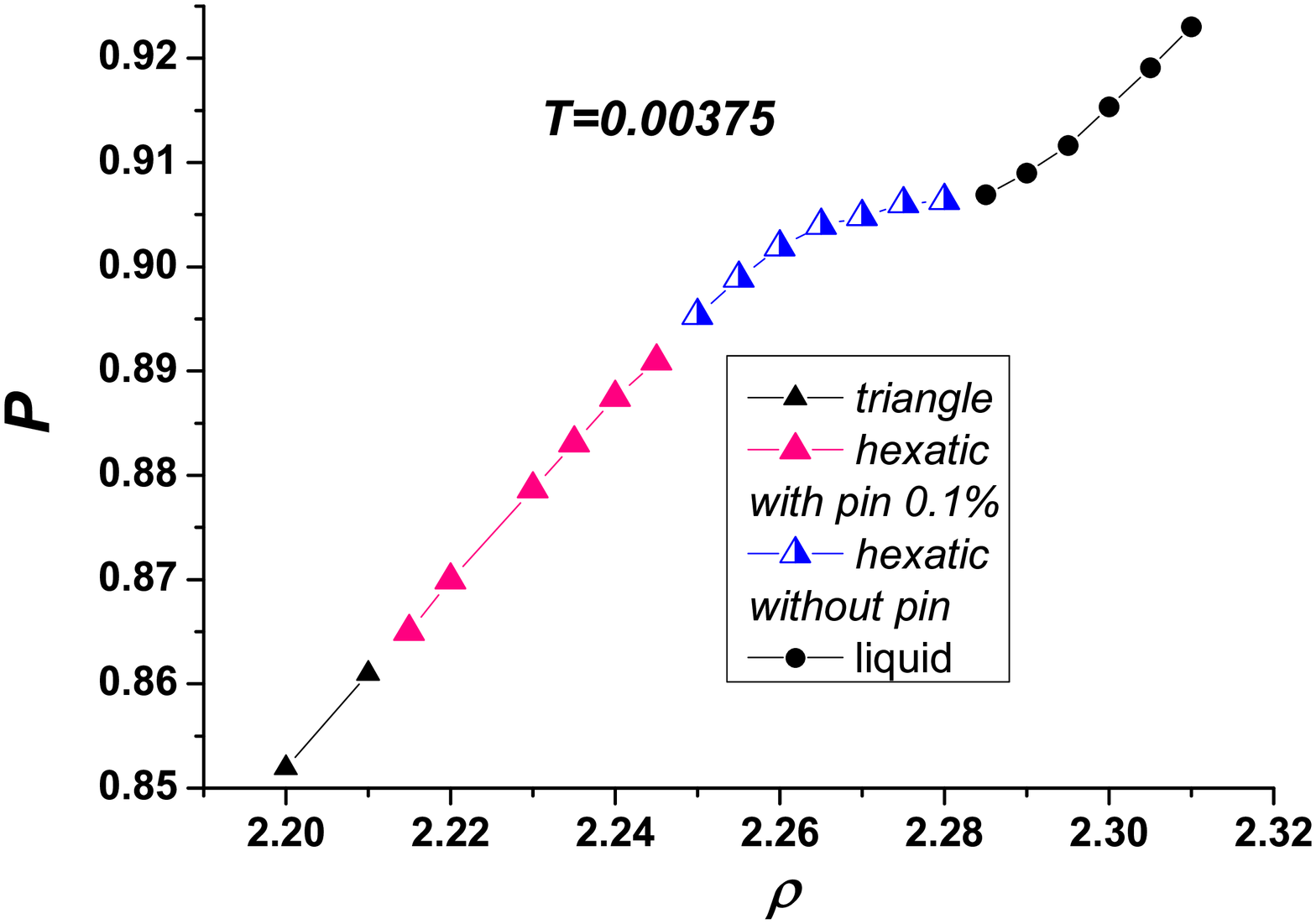}%

\includegraphics[width=8cm]{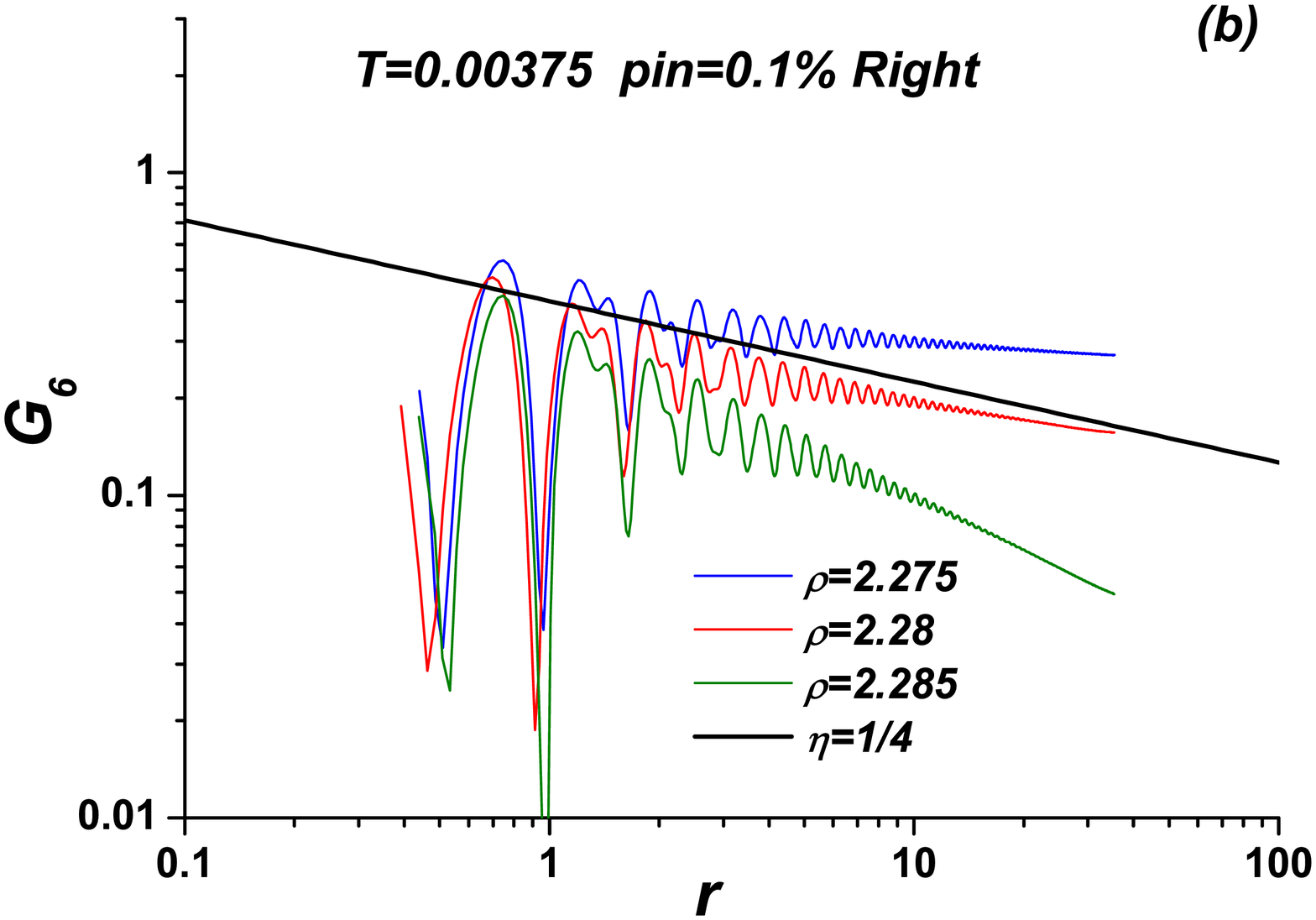}%

\includegraphics[width=8cm]{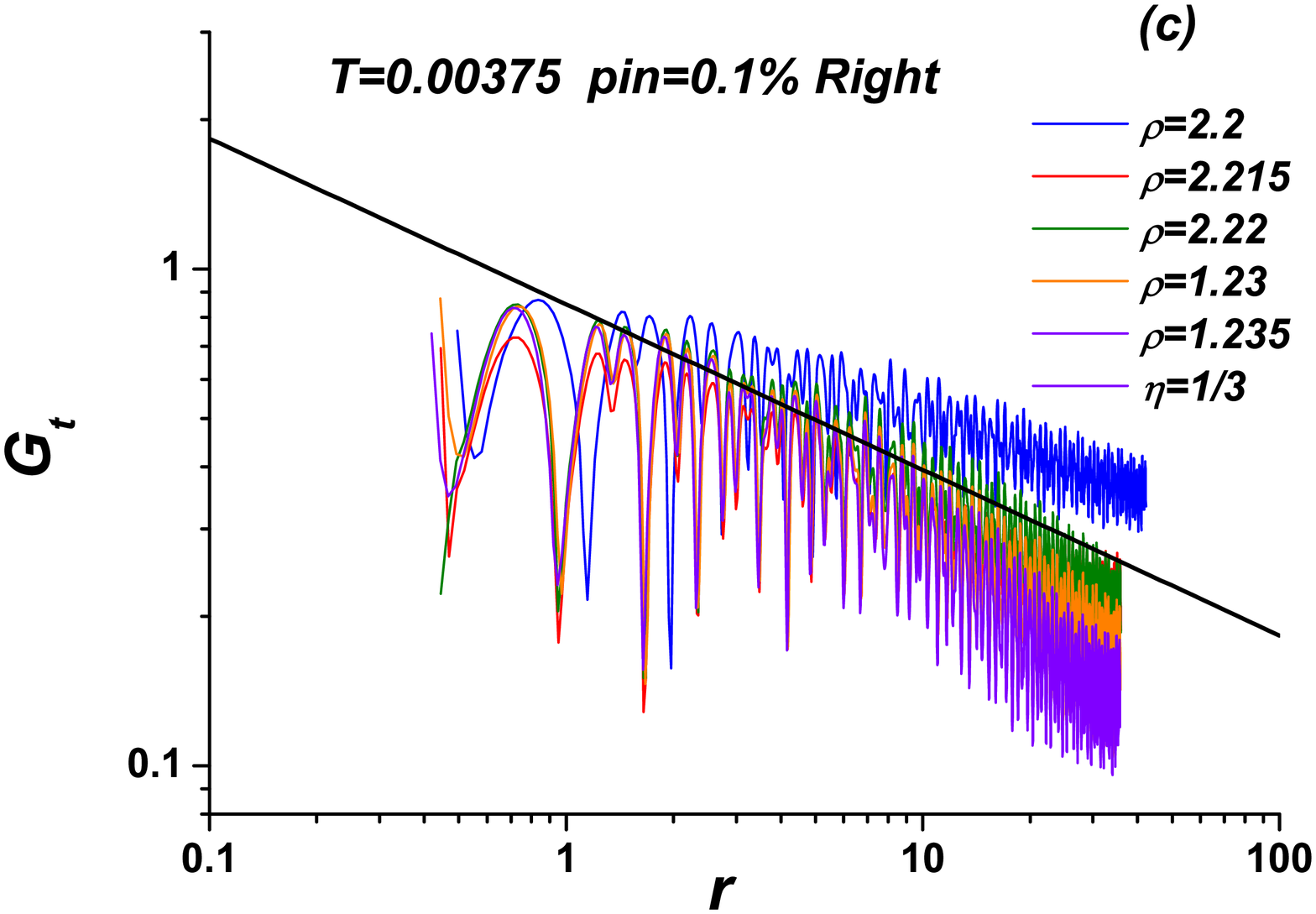}%

\caption{\label{tr-right} (a) The equation of state of Hertzian
spheres at $T=0.00375$ with concentration of random pinning $0.1
\%$ at the right branch of the melting line of the triangular
crystal. (b) The behavior of OCF $G_6$ of the same system. (c) The
behavior of TCF $G_t$ of the same system.}
\end{figure}

Fig.~\ref{tr-right} shows an equation of state without the
Mayer-Wood loop and the behavior of the correlation functions at
the crossing of the right branch of the triangular crystal melting
curve at $T=0.00375$ (see Fig.~\ref{pd-pin}). In this area in the
absence of random pinning the system melts via two continuous
transitions of the BKT type in accordance with the BKTHNY theory
\cite{molphys}. We can see that in this case the introduction of
random pinning also did not change the transition scenario but
only significantly expanded the hexatic phase existence area.

\section{The influence of random pinning on square crystal melting}

In this part of the paper we address the influence of random
pinning on melting of the square crystal in Hertzian spheres. In
order that the effect of random pinning become more vivid, we
increased the concentration of pinned particles to $0.2 \%$.
Searching for the distinct boundary between the square crystal and
the tetratic phase was carried out similarly to the procedure
described above for the triangular crystal.

\begin{figure}
\includegraphics[width=8cm]{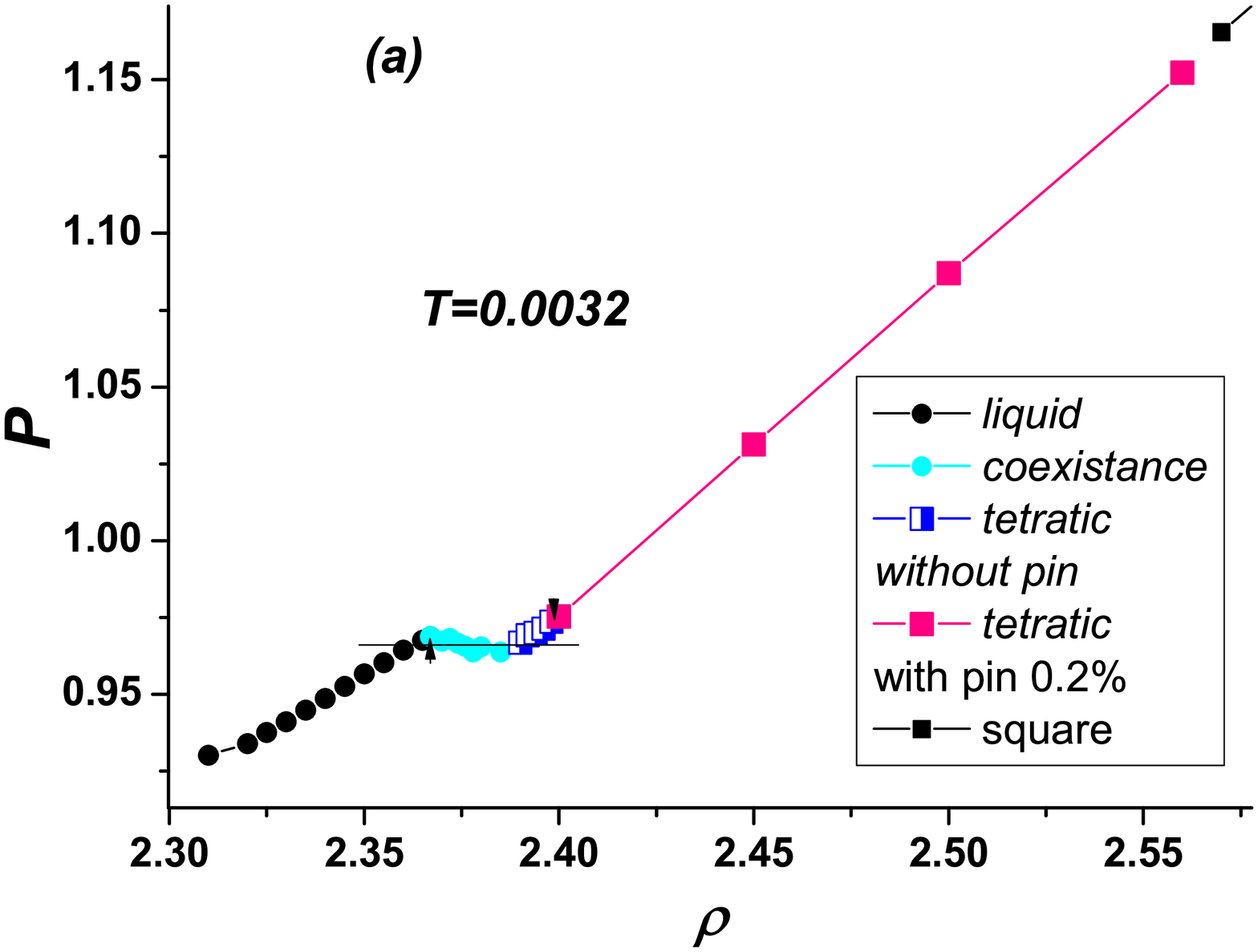}%

\includegraphics[width=8cm]{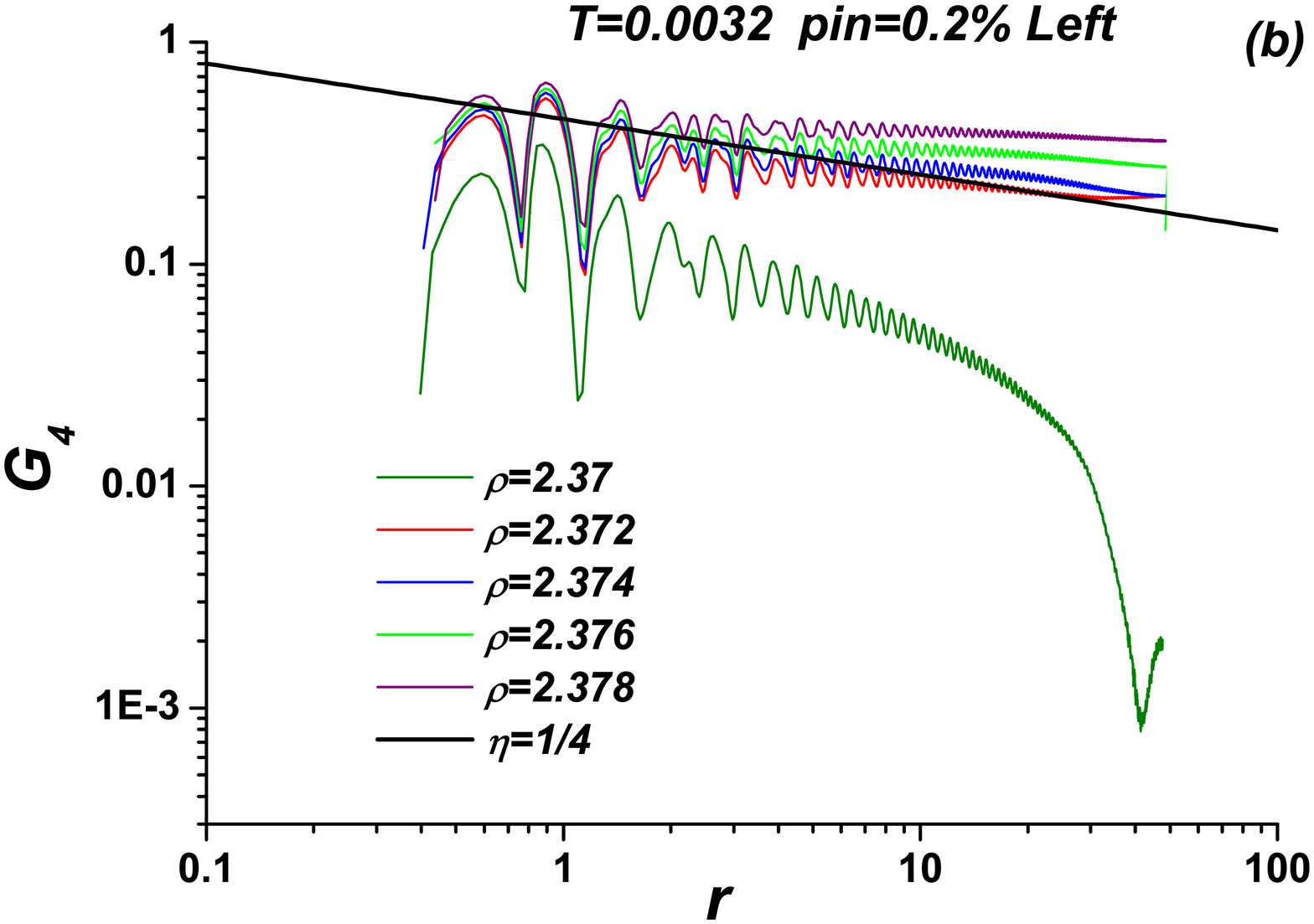}%

\includegraphics[width=8cm]{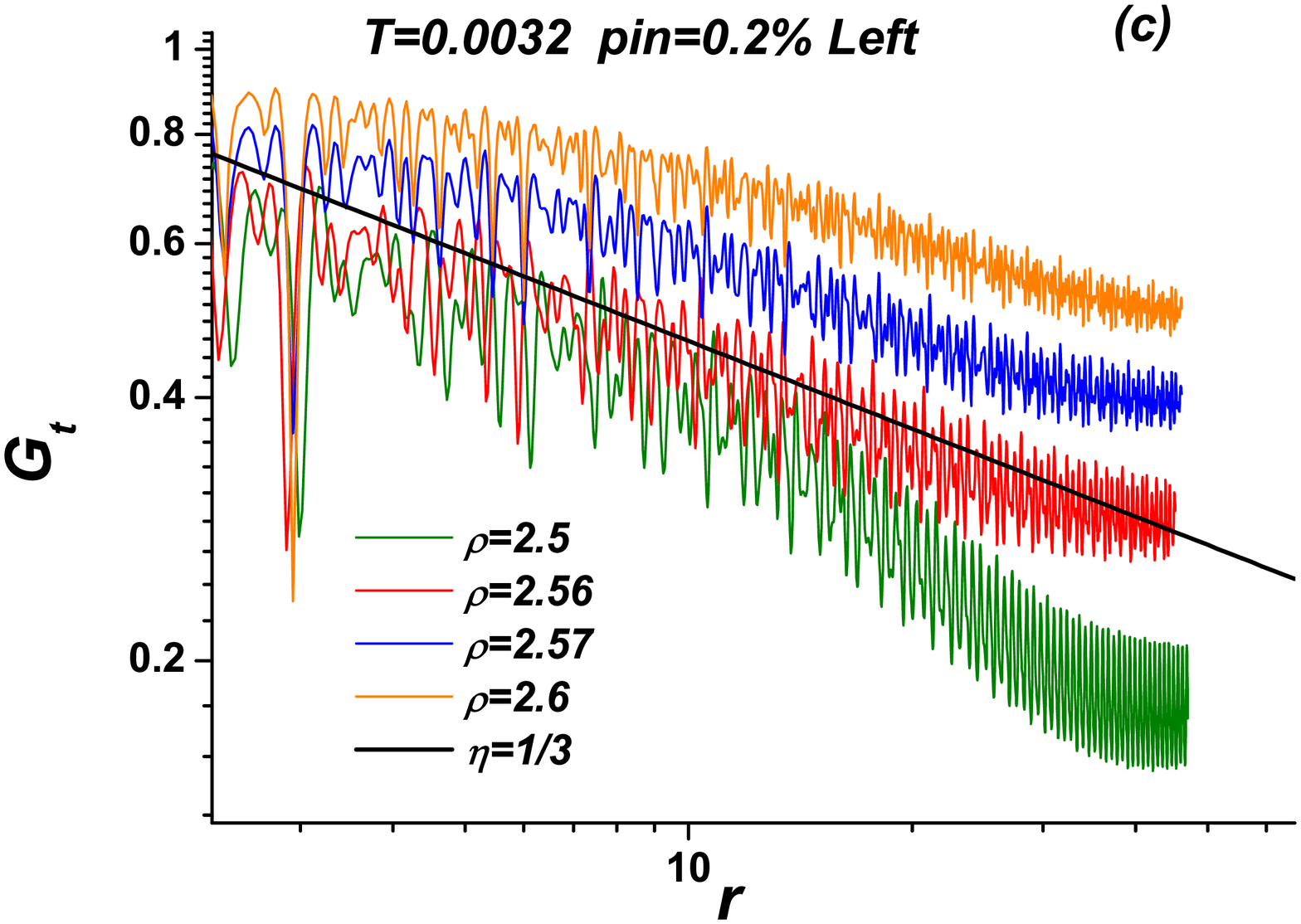}%

\caption{\label{sq-left} (a) The equation of state of Hertzian
spheres at $T=0.0032$ with concentration of random pinning $0.2$
\% at the left branch of the melting line of the square crystal;
(b) orientational correlation function $G_4$ at $T=0.0032$ and (c)
translational correlation function $G_t$ along the same isotherm.}
\end{figure}

Fig.~\ref{sq-left} (a) shows the equation of state of the system
with random pinning at $T=0.0032$ at the crossing of the left
branch of the square crystal melting line, on which the Mayer-Wood
loop was found, and in Fig.~\ref{sq-left} (b) and (c) - the
orientational and translational order parameter correlation
functions for the square crystal are shown. In order to visualize
the influence of random pinning on the tetratic phase we also show
in panel (a) the limit of stability of the square crystalline
phase without pinning that goes out of the Mayer-Wood loop. In
this region without random pinning the crystal to tetratic
transition is a continuous one of the BKT type, whereas the
tetratic to isotropic liquid transition is a first-order
transition \cite{molphys}. It is evident that in the presence of
random pinning the area of tetratic phase existence has
significantly expanded without changing the melting scenario,
which is in qualitative agreement with the influence of pinning on
triangular crystal melting for the third melting scenario.

\begin{figure}
\includegraphics[width=8cm]{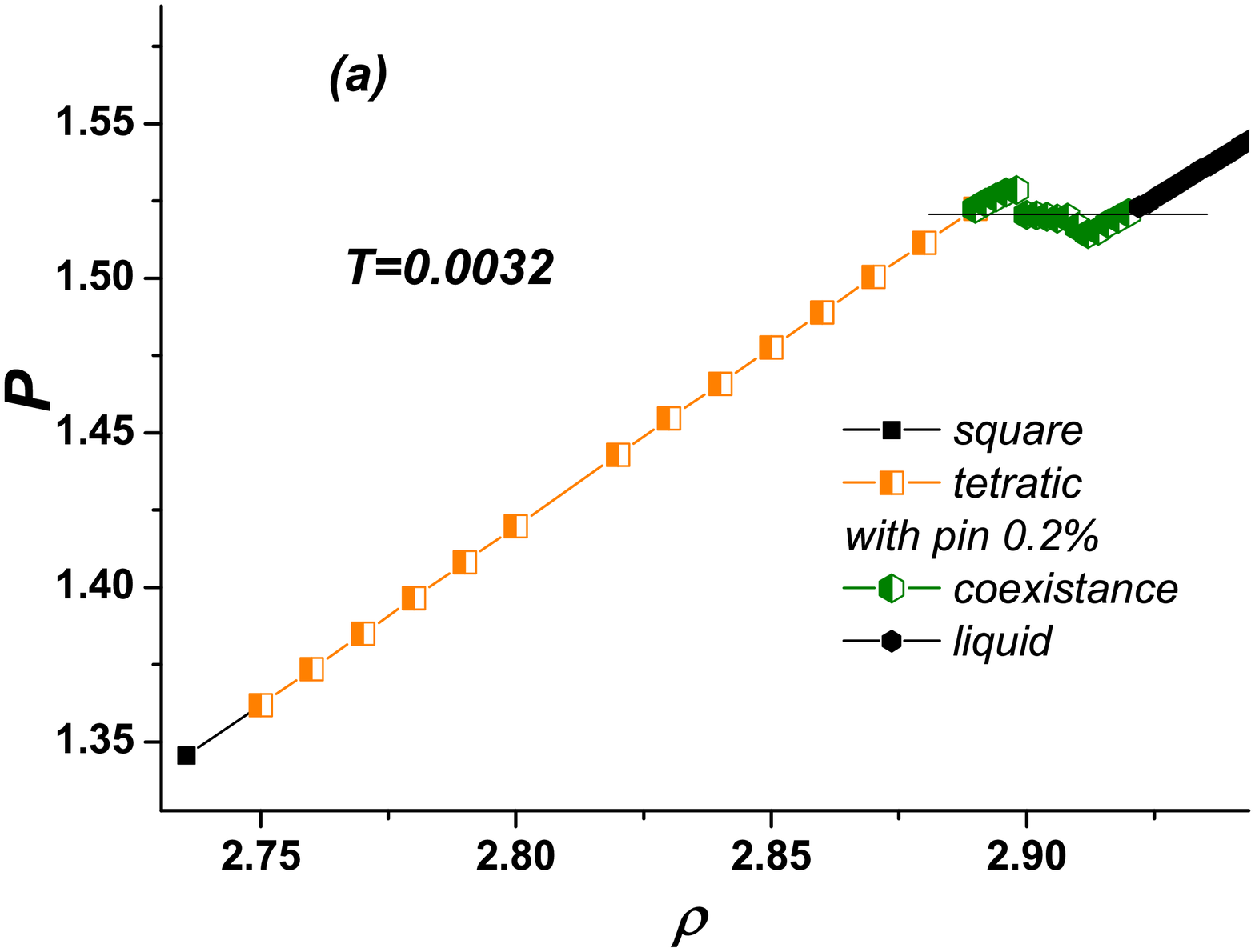}%

\includegraphics[width=8cm]{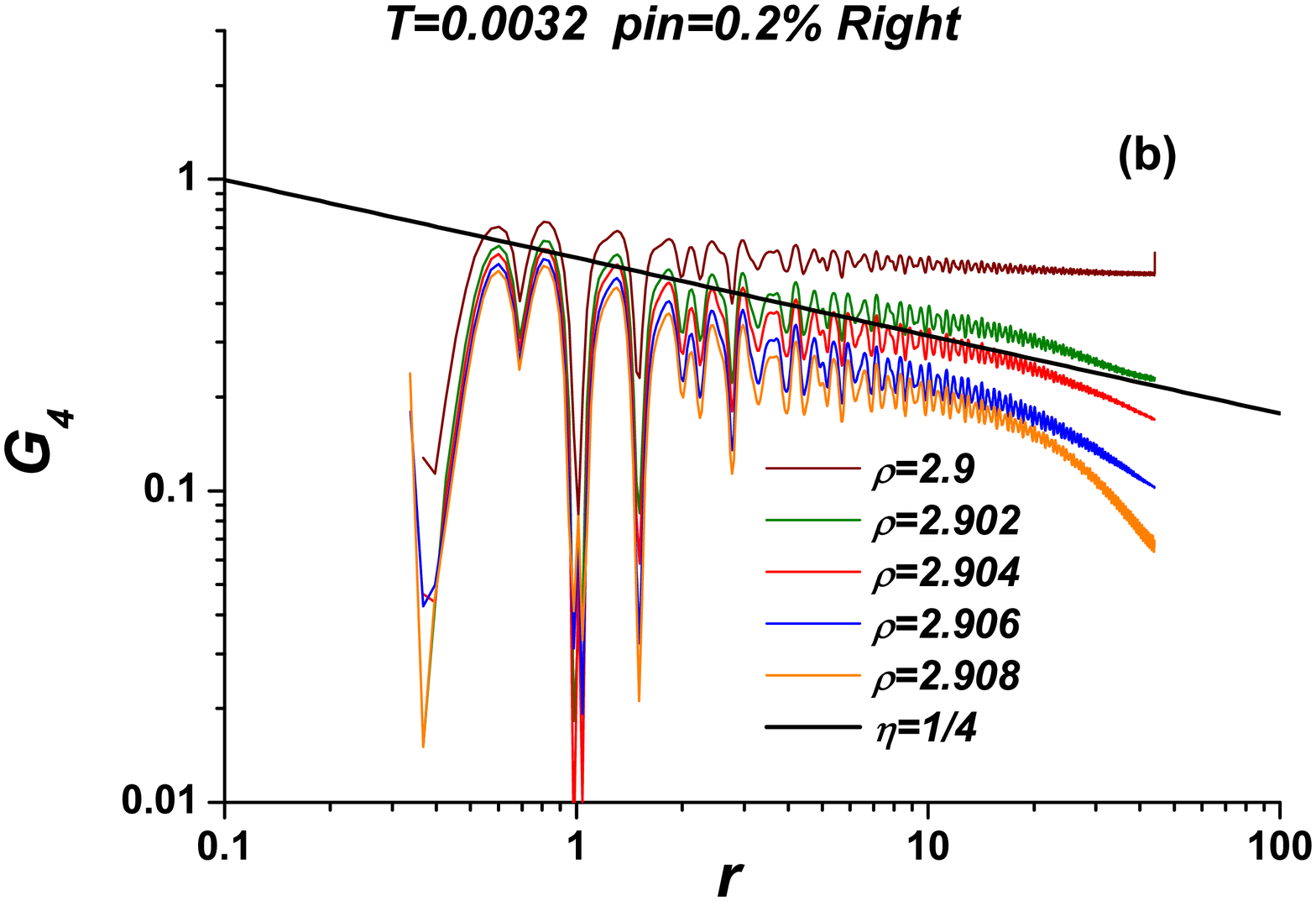}%

\includegraphics[width=8cm]{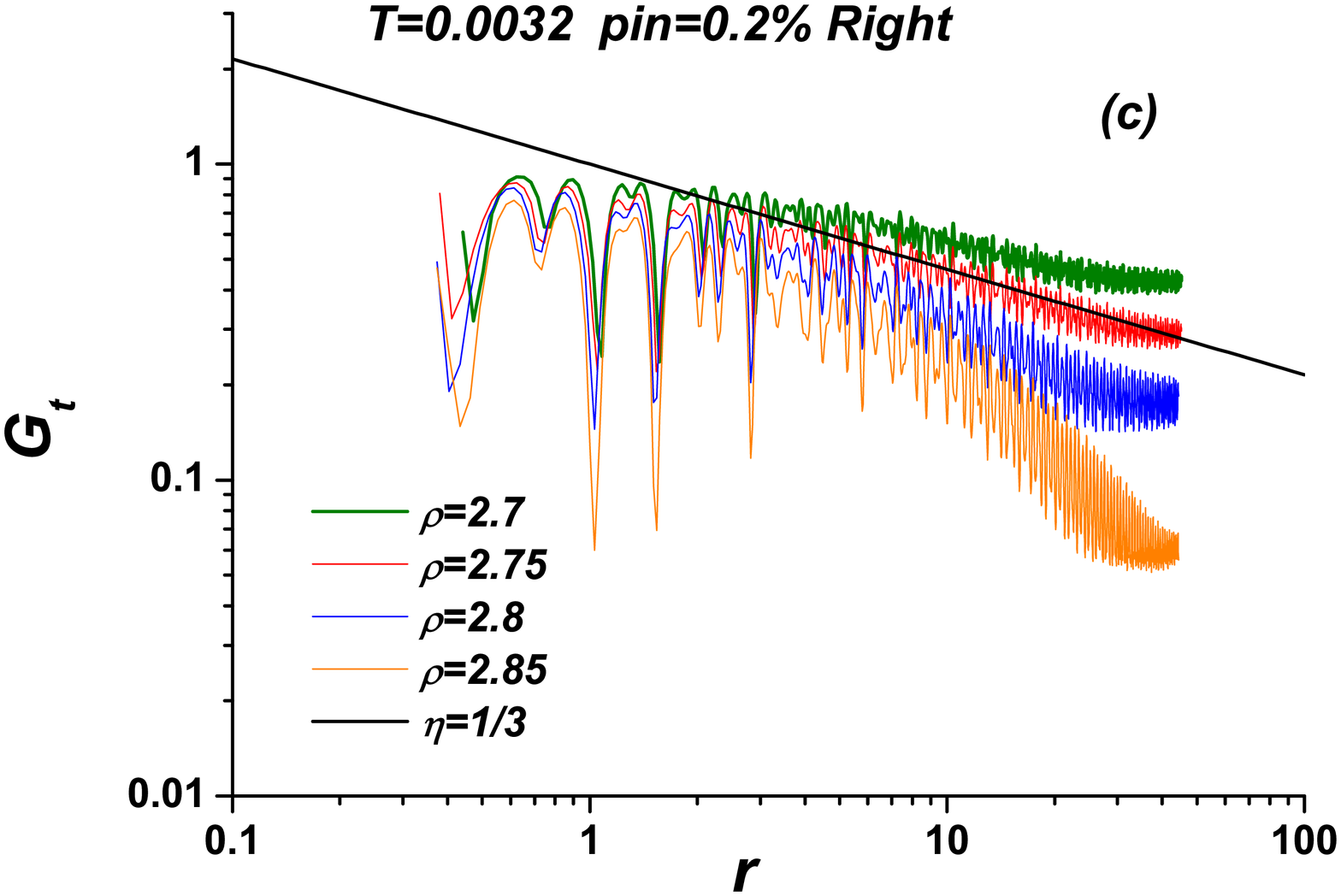}%

\caption{\label{sq-right} (a) The equation of state of Hertzian
spheres at $T=0.0032$ with concentration of random pinning $0.2$
\% at the right branch of the square crystal melting line; (b)
orientational correlation function $G_4$ at $T=0.0032$ and (c)
translational correlation function $G_t$ along the same isotherm.}
\end{figure}

In Fig.~\ref{sq-right} the same analysis is performed for the
right branch of the square crystal melting curve. In panel (a) the
equation of state with the Mayer-Wood loop at the crossing of the
melting curve right branch at $T=0.0032$ is shown. Again, we can
see a very wide area of the tetratic phase under the influence of
pinning, the border of which was determined from the behavior of
$G_t$ in Fig.~\ref{sq-right} (c). Recall that in the system
without random pinning this branch melted via one first-order
transition \cite{molphys}(both criteria following from the
behavior of $G_4$ and $G_t$ are inside the Mayer-Wood loop), i.e.,
in this case we observe a qualitative change in the melting
scenario: in the presence of random pinning the system melts
according to the third scenario.

Thus, the introduction of random pinning significantly influences
square phase melting. As in the triangular crystal case, a
considerable increase in the tetratic phase existence area occurs.
Moreover, if for the left branch the changes are limited to the
shifting of the crystal stability line then in the right one a
change in the crystal melting scenario takes place from one
first-order transition to the third type of melting.

\section{The influence of random pinning on transition from the triangular lattice to the square one}

In the preceding paragraphs it was shown that random pinning could
significantly affect 2D crystal melting. This influence is
primarily connected with a considerable growth of the existence
areas of the phases that are an intermediary between the crystal
and liquid (the hexatic phase for the triangular crystal and
tetratic for square). In some cases, this intermediary phase does
not exist at all in the system without pinning but appears in its
presence \cite{dfrt5}. This enables us to suppose that at low
temperatures in the neighborhood of transition between two
crystals the stability limit of the triangular phase will move to
lower densities and that of the square phase will be shifted to
higher densities. One can conclude that the structural transition
will take place not between two crystals but with the
participation of the hexatic and tetratic phases. This is a new
mechanism of structural transition in two dimensions in the
presence of random pinning. We believe that this mechanism is
general and independent of the form of the potential. This part of
the paper is devoted to verifying this mechanism. In doing so we
will simulate systems with random pinning concentration $0.2 \%$.

\begin{figure}
\includegraphics[width=8cm]{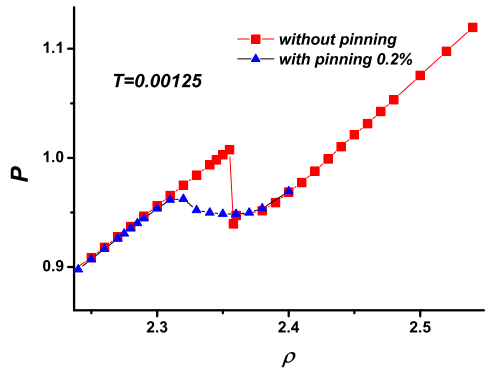}%

\caption{\label{tr-sq}  The equation of state of Hertzian spheres
at $T=0.00125$ with concentration of random pinning $0.2 \%$ at
the crossing of the first-order phase transition from the
triangular to square crystal as compared with the system without
pinning.}
\end{figure}

Fig.~\ref{tr-sq} shows the equation of state at $T=0.00125$ in the
area of densities that crosses the transition line from the
triangular crystal to the square one. For comparison, the results
for the system with and without random pinning are shown. It can
be seen that in the system without random pinning a sharp fall
from one phase to the other takes place in the middle of the
two-phase area. Such behavior, to all appearances, is connected
with the effects of metastability in transitioning between the two
crystals that can be resolved by means of lengthy simulation or by
the introduction of defects in the form of random pinning. As can
be seen from Fig.~\ref{tr-sq} in the system with pinning the
Mayer-Wood loop already has a smooth standard form.

\begin{figure}
\includegraphics[width=8cm]{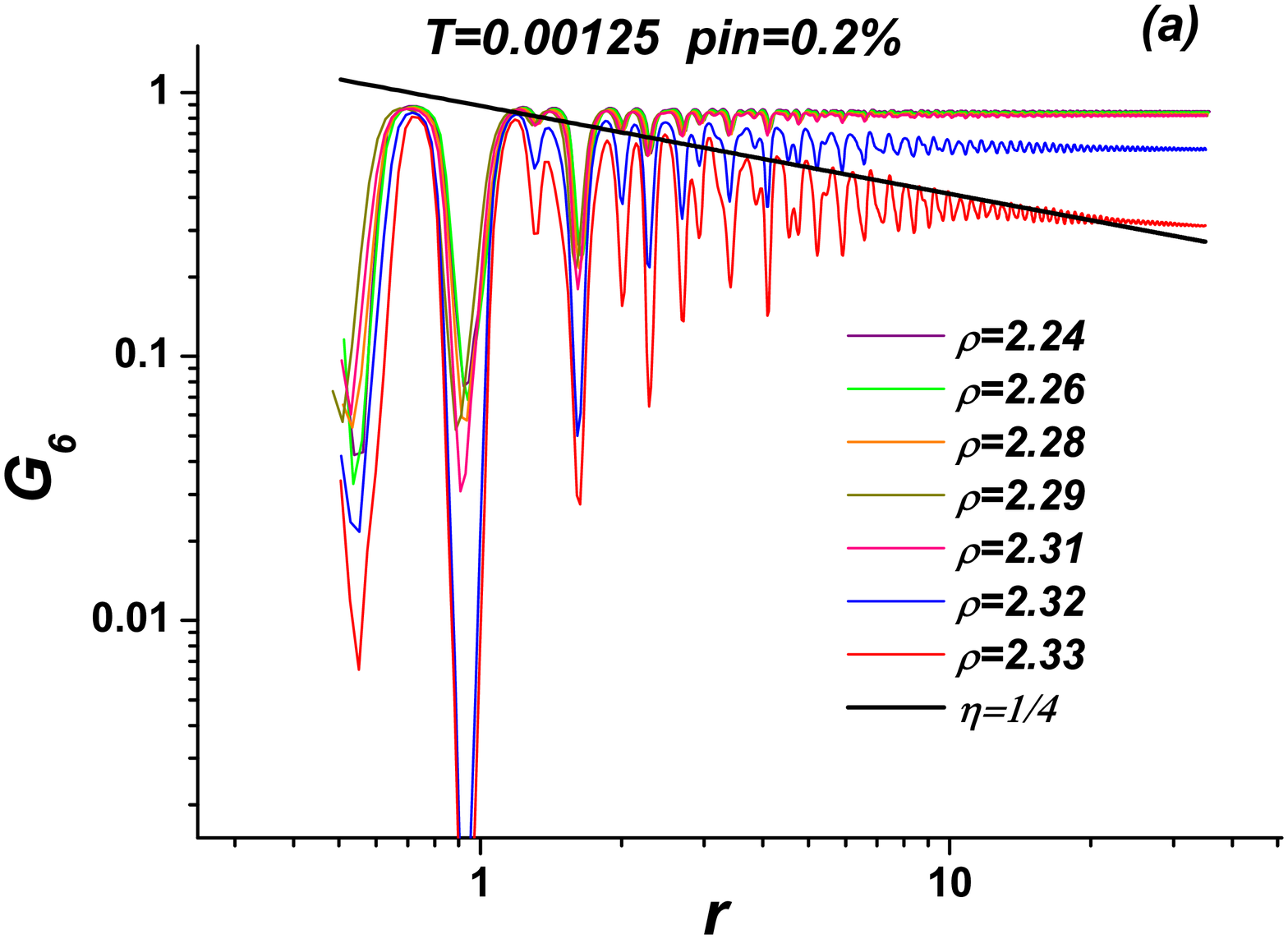}%

\includegraphics[width=8cm]{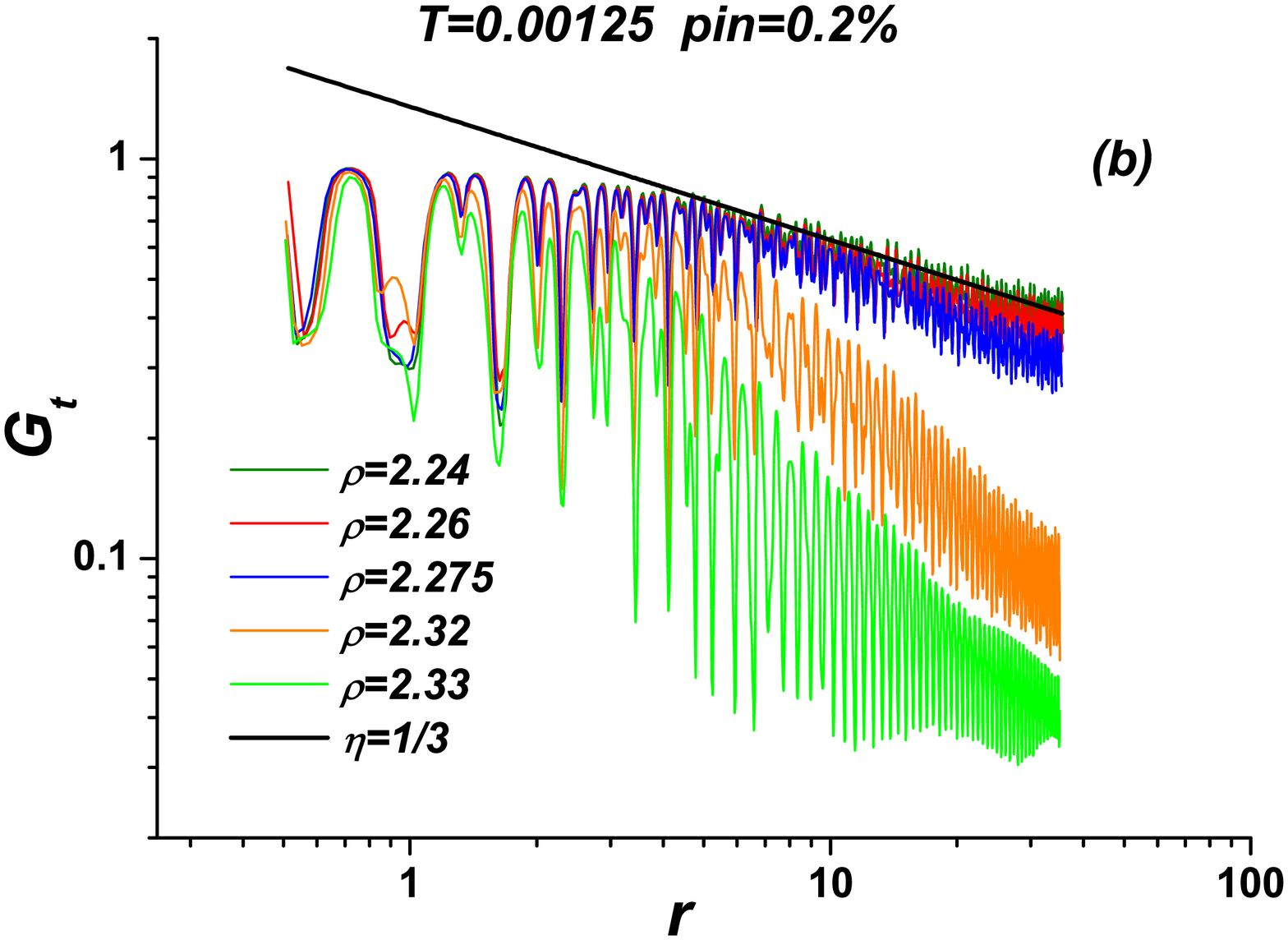}%

\caption{\label{tr-sq-left} (a) The orientational and (b)
translational correlation functions with the symmetry of a
triangular lattice with concentration of random pinning $0.2 \%$
in the region of transition from the triangular to square crystal
at $T=0.00125$.}
\end{figure}

\begin{figure}
\includegraphics[width=8cm]{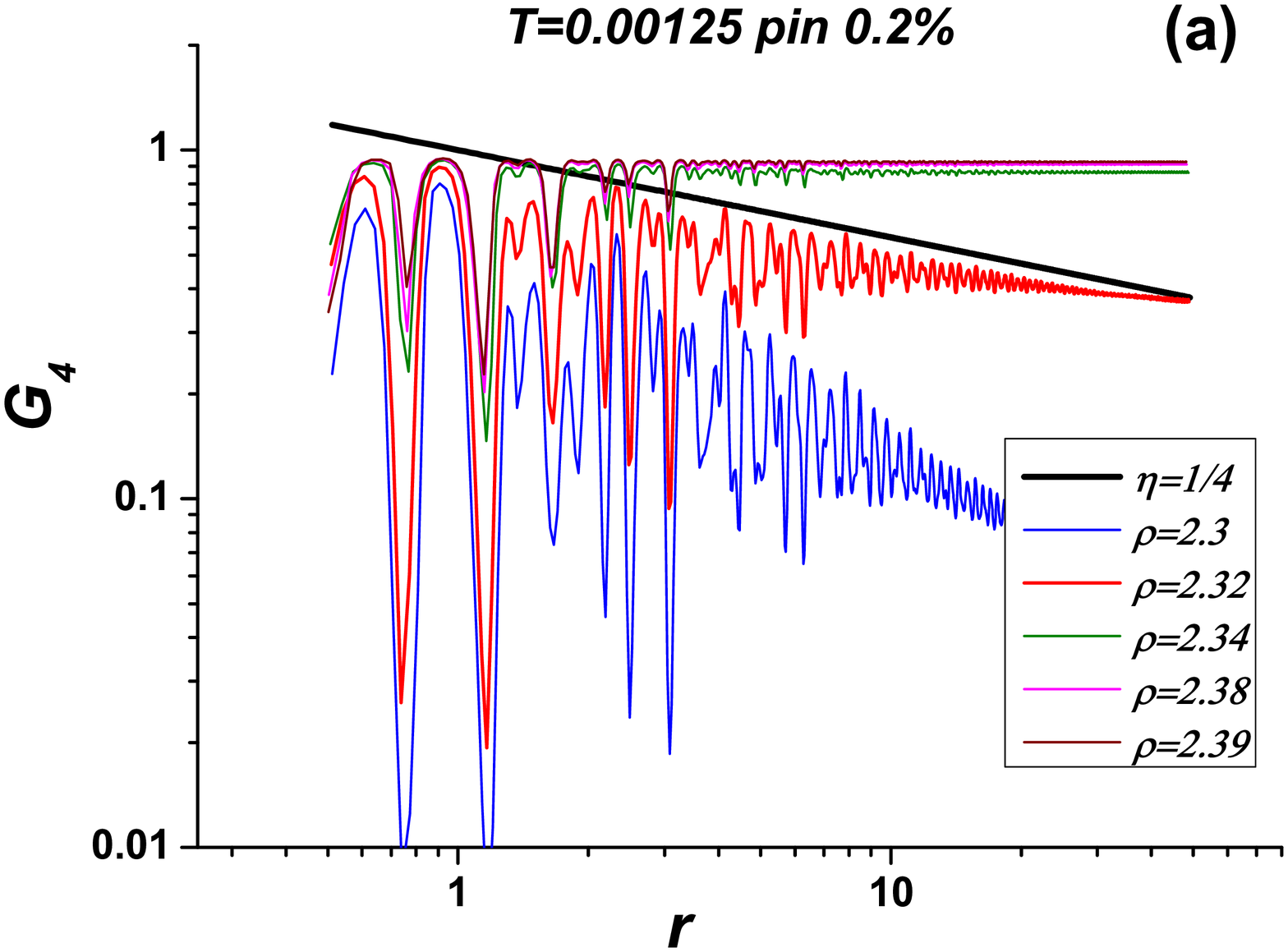}%

\includegraphics[width=8cm]{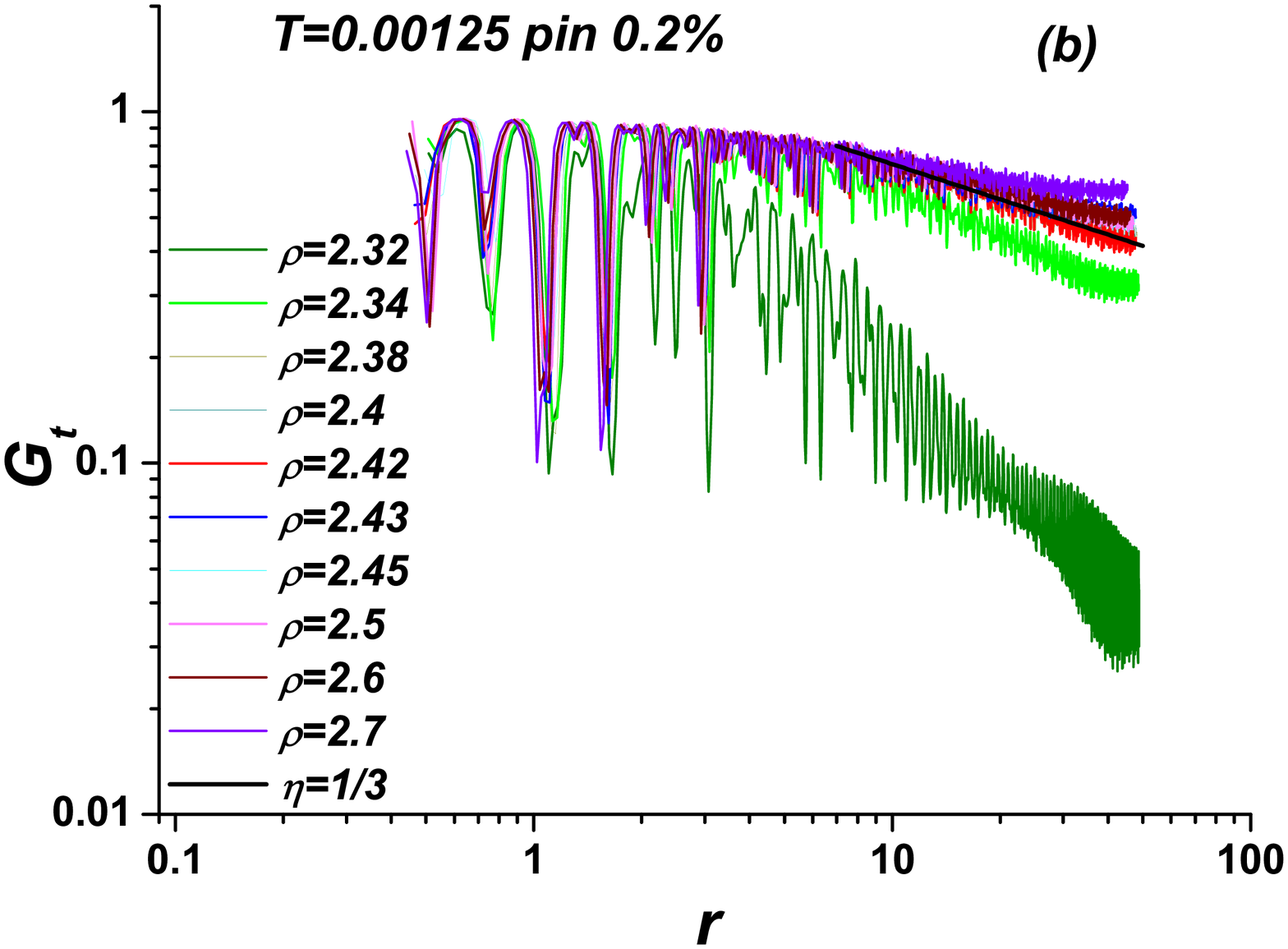}%

\caption{\label{tr-sq-right} (a) The orientational and (b)
translational correlation functions with the symmetry of a square
lattice with concentration of random pinning $0.2 \%$ in the
region of transition from the triangular to square crystal at
$T=0.00125$.}
\end{figure}

In order to understand the nature of change of the equation of
state along isotherm $T=0.00125$ under the influence of pinning
(Fig. \ref{tr-sq}) in the metastable area (for instance, to the
coexistence of what phases will it correspond?) and beyond it let
us consider the correlation functions of translational and
orientational order for the triangular lattice to the left of the
Mayer-Wood loop and inside of it; and do the same for the square
one to the right of the loop and inside of it, respectively. Fig.
\ref{tr-sq-left} shows the correlation functions corresponding to
the triangular lattice. From these figures it is evident that at
density $\rho=2.26$ the triangular crystal loses stability and
continually transitions according to BKT to the hexatic phase,
which in turn, loses stability at $\rho=2.33$ that is inside the
loop. An analogous situation arises for the square crystal: from
the correlation functions in Fig. \ref{tr-sq-right} it is evident
that the tetratic stability limit is located inside the loop at
$\rho=2.32$, while the stability limit of the crystal itself is
outside of it at $\rho=2.42$, which is evidence of its continuous
transition of the BKT type to the tetratic phase. Consequently,
the loop itself corresponds to the hexatic-tetratic coexistence
area and to the first-order transition between these phases owing
to the basic difference of their orientational symmetry. With a
further increase in density a continuous BKT transition from the
tetratic phase to the square crystal takes place.

\begin{figure}
\includegraphics[width=8cm]{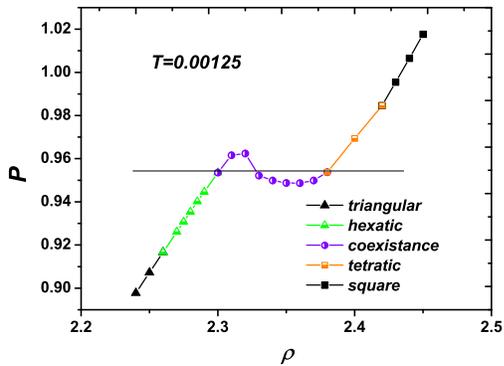}%
\caption{\label{tr-sq-color} The sequence of phases in transition
from the triangular to square crystal in the system with 0.2 \% of
random pinning at $T=0.00125$.}
\end{figure}

Thus in the presence of random pinning the scenario of transition
between the two crystals changes fundamentally, in this case
between the triangular and square crystals. Whereas in the system
without random pinning this transition occurs via a first-order
phase transition, in the presence of random pinning transformation
from the triangular crystal into the square one takes place via a
whole cascade of transitions of different nature. As it was shown
previously \cite{dfrt5}, random pinning could transform a
first-order melting transition into two transitions corresponding
to the third melting scenario. In the present case, one
first-order structural transition changed into three transitions.
First, the triangular crystal continuously transforms into the
hexatic phase. After that, the hexatic phase transitions to
tetratic via a first-order phase transition. Finally, the tetratic
phase continuously transforms into the square crystal. This
scenario is illustrated in Fig. \ref{tr-sq-color}.

It is noteworthy that the mechanism of transition between crystals
detected by us in the presence of random pinning is in qualitative
agreement with the influence of random pinning on crystal melting
\cite{dfrt5}. Random pinning destabilizes the crystal and expands
the existence area of the hexatic (in the case of the square
crystal, tetratic) phase. Moreover, the introduction of random
pinning can generate the hexatic or tetratic phase if in the
system without it melting took place by means of first-order
transition without the hexatic or tetratic phase. An example of
such behavior is the right branch of square crystal melting in the
system under investigation. Similarly, in transition from one
crystal to another random pinning destabilizes the crystal before
reaching the first-order transition line (the Mayer-Wood loop) and
transforms it into the hexatic (tetratic for the square crystal)
phase.

\begin{figure}
\includegraphics[width=8cm]{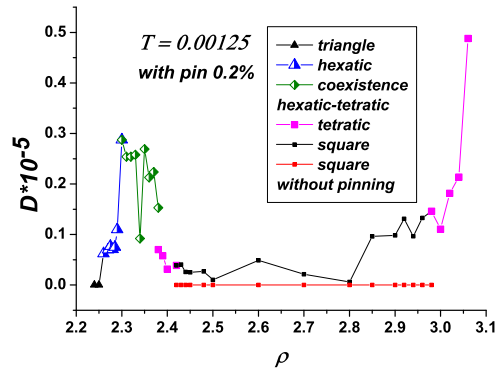}%

\caption{\label{dif-tr-sq} Diffusion coefficient behavior
depending on density in the system with pinning $0.2 \%$ and in
the transition area between the triangular and square crystals at
$T=0.00125$.}
\end{figure}

Fig. \ref{pd-pin} represents the phase diagram of Hertzian spheres
both without random pinning and with random pinning with
concentrations $0.1 \%$ and $0.2 \%$, which reflects the given
results and considerations. It should be noted that while in order
to watch the influence of random pinning on the melting curve it
was sufficient to add only $0.1 \%$ of fixed particles, to observe
the influence of random pinning on transition between the crystals
it was necessary to increase the concentration of fixed particles
to $0.2 \%$. This may be due to the fact that the overall mobility
of the particles during transition from one crystal to another is
lower than during melting of the system. To illustrate this, let
us consider the behavior of the diffusion coefficient at
$T=0.00125$ in the density range from $\rho =2.24$ to $3.06$ (Fig.
\ref{dif-tr-sq}). The lowest densities of this range correspond to
the triangular crystal. The highest - to the tetratic phase at the
densities up to square crystal melting on the right branch. The
values of the diffusion coefficient in the different phases are
denoted by different symbols. As expected, in the crystals without
random pinning the diffusion coefficient is near zero. On the
introduction of random pinning, the diffusion coefficient becomes
somewhat above zero, which is characteristic of crystals with
defects. The diffusion coefficient sharply increases in the
hexatic and tetratic phases as well as in the area of their
coexistence.

\begin{figure}
\includegraphics[width=8cm]{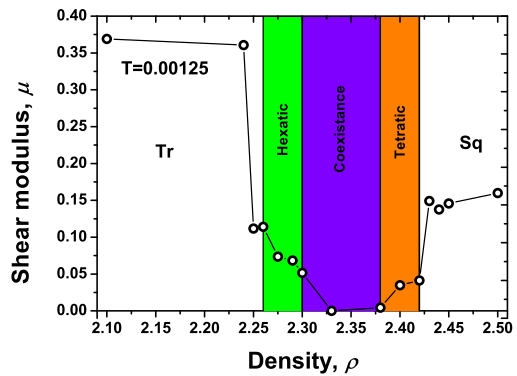}%

\caption{\label{mu-tr-sq} The shear modulus depending on density
in the system with pinning in the transition area between the
triangular and square crystals at $T=0.00125$.}
\end{figure}

Similar conclusions can be made from shear modulus behavior shown
in Fig. \ref{mu-tr-sq}. It is evident that the shear modulus has
great importance in the crystalline phases. In transition to
hexatic or tetratic the shear modulus falls sharply. Importantly,
in the presented simulation the shear modulus in the hexatic and
tetratic phases turns out to be more than zero, which is due to
simulation limitations. However, it might become strictly zero
after solving the renormalization group equations shown in the
Appendix. In the coexistence area of hexatic and tetratic the
shear modulus becomes zero.

\section{Conclusions}

We study the influence of random pinning on the phase diagram of
2D Hertzian spheres with $\alpha =5/2$. The full phase diagram of
a system with pinning in the existence area of the triangular and
square crystals was built. It was shown that random pinning
decreased the stability area of crystalline phases and increased
the area of stability of the hexatic and tetratic phase as well as
changing the melting scenario of the square crystal from one
first-order transition to the third type of melting, which is in
qualitative agreement with the previous works. For the first time,
the influence of random pinning on transition between two
crystalline phases was investigated. It was shown that the
introduction of random pinning significantly changed the mechanism
of transition. While in the system without random pinning
transition took place as a first-order transition, the
introduction of pinning makes it a three-stage one: the triangular
crystal continuously transforms into hexatic, the hexatic via a
first-order transition transforms into tetratic, after which the
tetratic continuously transforms into the square crystal. This
mechanism is in qualitative agreement with predictions on the
influence of random pinning on the melting curve and may, in a
certain sense, be considered as an extension of the BKTHNY melting
mechanism to transition between two 2D crystalline phases.

\begin{acknowledgments}
This work was carried out using computing resources of the federal
collective usage center "Complex for simulation and data
processing for mega-science facilities" at NRC "Kurchatov
Institute", http://ckp.nrcki.ru, and supercomputers at Joint
Supercomputer Center of the Russian Academy of Sciences (JSCC
RAS). The work was supported by the Russian Science
Foundation (Grants No 19-12-00092).
\end{acknowledgments}

\appendix
\section{}

\begin{figure}
\includegraphics[width=8cm]{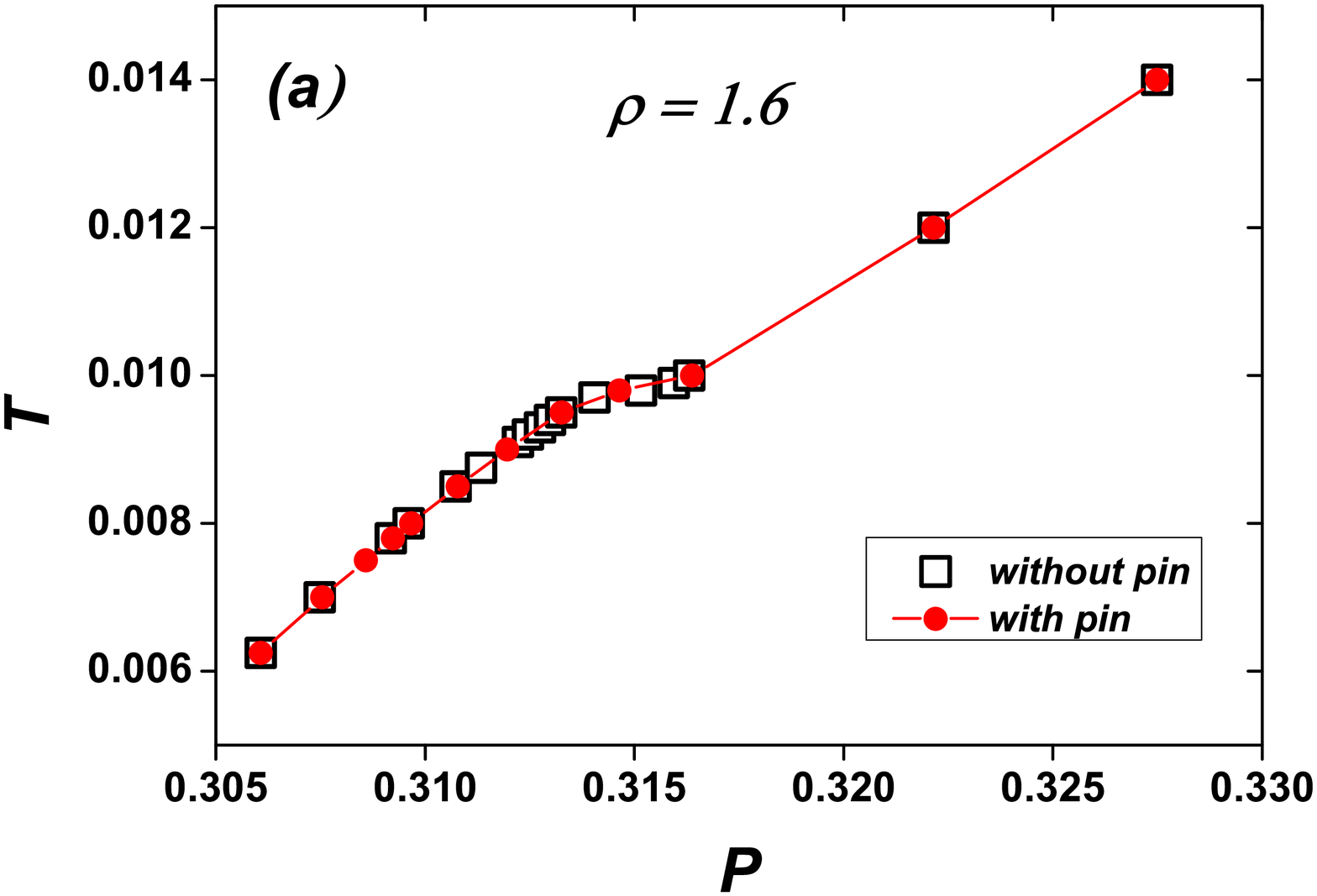}%

\includegraphics[width=8cm]{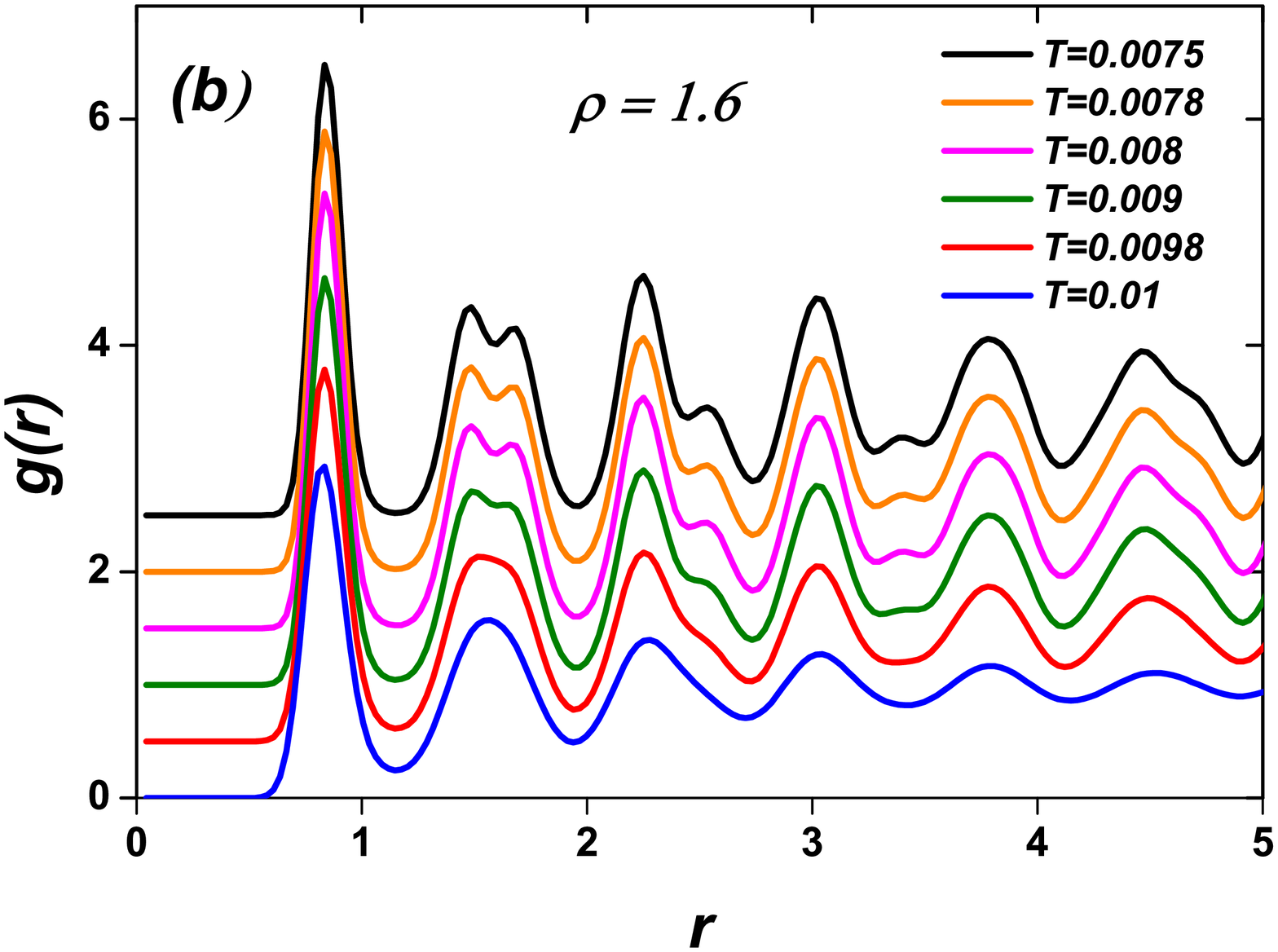}%

\caption{\label{isoch-rdf} (a) Isochores $\rho=1.6$ for the system
without pinning and in the presence of pinning $0.1 \%$. (b)
Radial distribution functions $g(r)$ for the system with pinning
$0.1 \%$ as they approach the tricritical point. For better
visualization, the curves are shifted along axis $Y$ with a step
of 0.5.}
\end{figure}

Of special interest is the influence of random pinning on the
behavior of the triangular crystal melting curve near its maximum
which in the system without pinning takes place at density
$\rho=1.6$ \cite{molphys}. As it is seen in Fig.~\ref{pd-pin},
without pinning the width of the hexatic phase tends to zero in
the vicinity of the melting curve maximum. The point with maximum
temperature $T_{max}$ on the melting curve is a tricritical point
in which the hexatic-liquid transition undergoes a change in the
transition scenario from first-order at $\rho<1.6$ to continuous
of the BKT type at $\rho>1.6$. Can random pinning affect the value
of $T_{max}$? In Fig.~\ref{isoch-rdf} (a) isochores $\rho=1.6$ are
represented that, at first sight, fully coincide for the systems
without and with pinning $0.1 \%$. On both isochores the
inflection point at $T=0.0098$ is clearly seen that corresponds to
the tricritical point in the system without pinning
\cite{hertzmelt, molphys} and, as we suppose, in the system with
pinning it will have the same value. Fig.~\ref{isoch-rdf}(b) shows
the behavior of radial distribution functions $g(r)$ along
isochore $\rho=1.6$ for the system with pinning. From the figure
it can be seen that with an increase in temperature the splitting
of the second peak characteristic of the triangular crystal is
blurred forming a single peak and at $T=0.01$ the form of $g(r)$
corresponds to liquid. It can be supposed that the transition from
hexatic to liquid based on the behavior criterion of $g(r)$ occurs
at $T=0.0098$, which is in full agreement with the data on the
isochores.

\begin{figure}
\includegraphics[width=8cm]{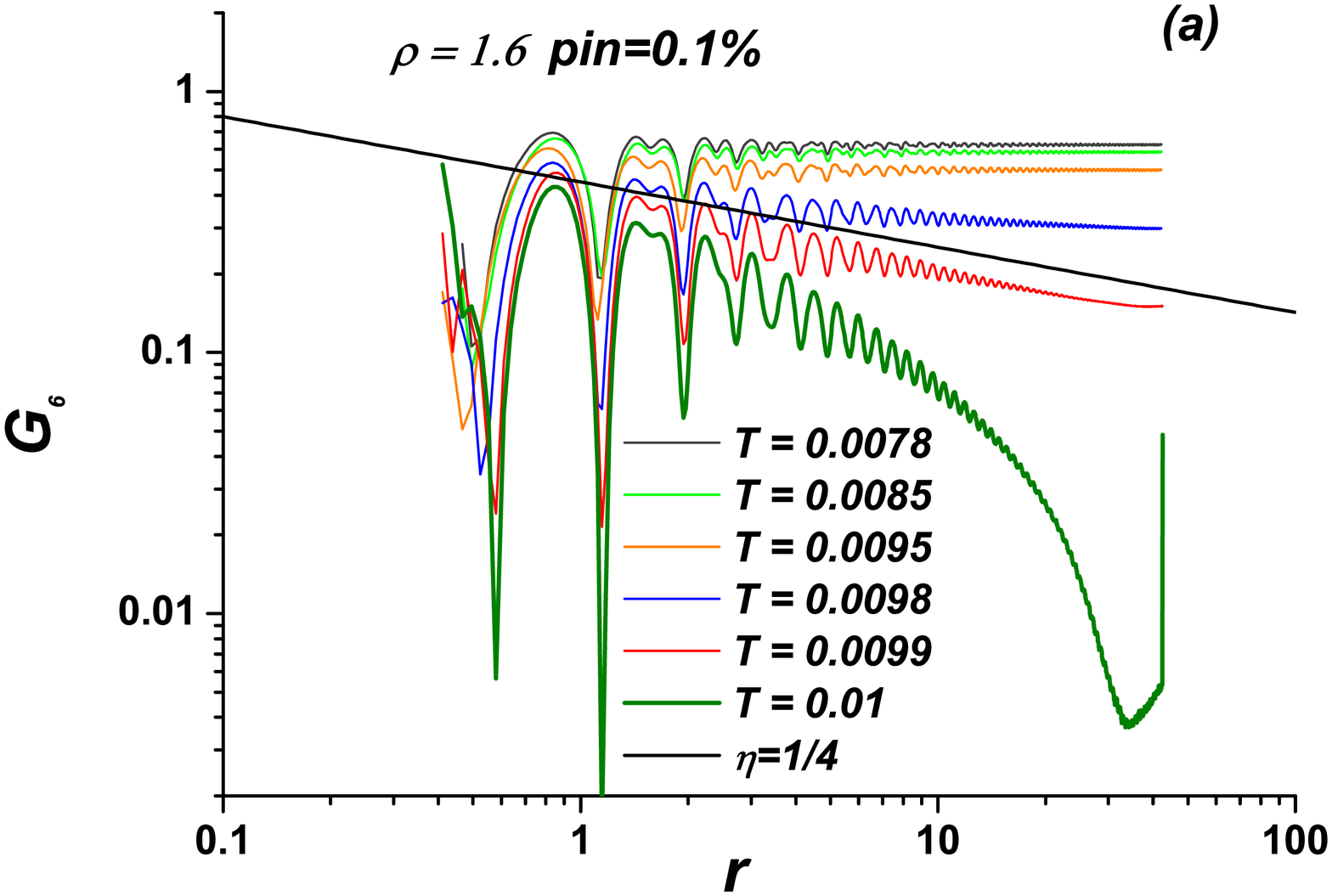}%

\includegraphics[width=8cm]{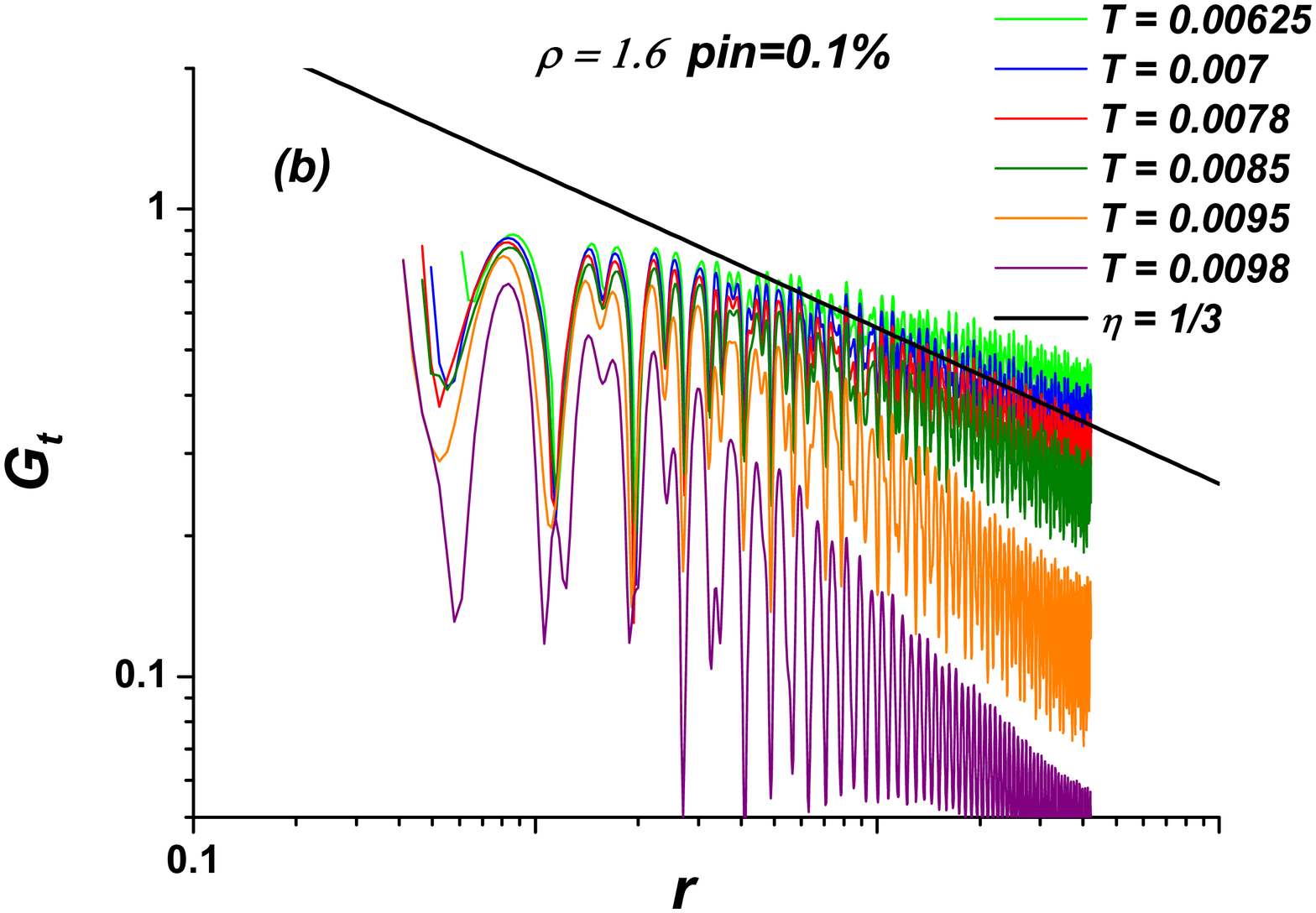}%

\caption{\label{tr-r16} (a) Orientational and (b) translational
correlation functions $G_6$ and $G_t$ along isochore $\rho =1.6$
at random pinning concentration $0.1 \%$.}

\end{figure}

However, these results do not provide information on the
crystal-hexatic transition boundary. In order to determine the
exact crystal-hexatic-liquid transition boundary we made use of
the criterion on the basis of studying the orientational and
translational order parameter correlation functions on isochore
$\rho=1.6$ at concentration of random pinning $0.1 \%$ which are
represented in Fig.~\ref{tr-r16}. It can be seen that the crystal
loses stability with respect to transition to hexatic at slightly
above $T=0.0078$, i.e., long before reaching the temperature of a
maximum ($T_{max}=0.0098$). Thus, the crystal finds itself
completely surrounded by the hexatic phase existence area. All
three criteria with high accuracy yielded the value of the
tricritical point at $T_{max}=0.0098$. This result emphasizes one
more time that random pinning does not affect the orientational
order parameter and its correlation function and, accordingly, the
hexatic phase stability limit.

\begin{figure}
\includegraphics[width=8cm]{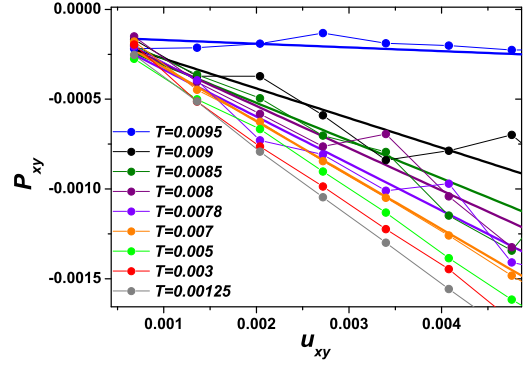}%

\caption{\label{mu-r16} The nondiagonal component of pressure as a
function of applied strain along isochore $\rho =1.6$ at random
pinning concentration $0.1 \%$.}
\end{figure}

In order to additionally confirm a substantial decrease in
crystal-hexatic transition temperature under the influence of
pinning at density $\rho=1.6$ we made use of the calculation and
the renormalization procedure of the Young modulus presented in
the works \cite{ray, Lutsko, Voyiatzis,Dijkstra2}. The elastic
properties of a triangular crystalline lattice may be fully
described by two independent elastic constants, namely, bulk
modulus $B$ and shear modulus $\mu$. Shear modulus $\mu$ is
calculated using the method suggested in \cite{broughton}. In this
method, the system is considered as strained. As a result, a
nondiagonal pressure component appears that is proportional to the
shear modulus:
\begin{equation}
  P_{xy}=\mu u_{xy}+O(u_{xy}^2),
\end{equation}
where $u_{xy}$ is strain. Fig.~\ref{mu-r16} shows $P_{xy}$
depending on the strain. It can be seen that in the crystal at low
temperatures far from the point of transition to the hexatic phase
the dependence is characterized by strong linearity. The accuracy
of calculations becomes worse when approaching transition to the
hexatic phase.

Young's modulus $K$ of a 2D triangular crystal that is a
combination of bulk modulus $B$ and shear modulus $\mu$ is
calculated from the relation
\begin{equation}
K=\frac{8}{\sqrt3 \rho k_BT} \frac{\mu (\mu + \lambda)}{2 \mu + \lambda},
\end{equation}
where Lame coefficient $\lambda$ is connected with bulk modulus $B
= (dP/d\rho)_T$ = $\mu$ + $\lambda$ \cite{kostth, halpnel2,
young}. The unrenormalized elastic constants and Young's modulus
only have sense for an ideal defect-free triangular lattice. In
the case of the presence of topological defects such as
dislocations, the renormalization procedure is simply obligatory
in accordance with the BKTHNY theory about a considerable decrease
in elasticity in the presence of dislocations.

In order to renormalize Young's modulus, first of all, we
determine dislocation core energy $E_c$ \cite{binder} that is
directly connected with the probability of detection of a
dislocation pair:
\begin{equation}
p_d=\frac{16 \sqrt3 \pi^2}{K-8 \pi} I_0(\frac{K(l)}{8 \pi})e^{\frac{K(l)}{8 \pi}}e^{-\frac{2E_c}{k_BT}},
\end{equation}
where $ I_0 $ and $ I_1 $ are modified Bessel functions
\cite{binder, fisher}. Note that $p_d = n_{dp} / N$, where
$n_{dp}$ is the number of dislocation pairs per the number of
particles $N$. We renormalize Young's modulus and fugacity of
dislocations $y$ using the recursive equations \cite{halpnel2,
young}:
\begin{equation}
\frac{d K^{-1}(l)}{dl}=\frac{3 \pi}{4} y^2(l)e^{\frac{K(l)}{8 \pi}} \left( 2I_0(\frac{K(l)}{8 \pi}) - I_1(\frac{K(l)}{8 \pi}) \right), \label{rg1}
\end{equation}

\begin{equation}
\frac{d y(l)}{dl}=\left( 2 - \frac{K(l)}{8 \pi} \right)y(l)+2 \pi y^2(l) e^{\frac{K(l)}{16 \pi}} I_0(\frac{K(l)}{8 \pi}), \label{rg2}
\end{equation}
where $l$ is the flowing number of renormalization group analysis.
The limit of an infinite system corresponds to infinitely large
$l$. Unrenormalized Young's modulus $K$ ($l$ = 0) and $y$ ($l$ =
0) = $\exp (-E_c/ k_BT)$ serve as initial conditions for connected
differential equations (Eqns. (\ref{rg1}) and (\ref{rg2})).

\begin{figure}
\includegraphics[width=8cm]{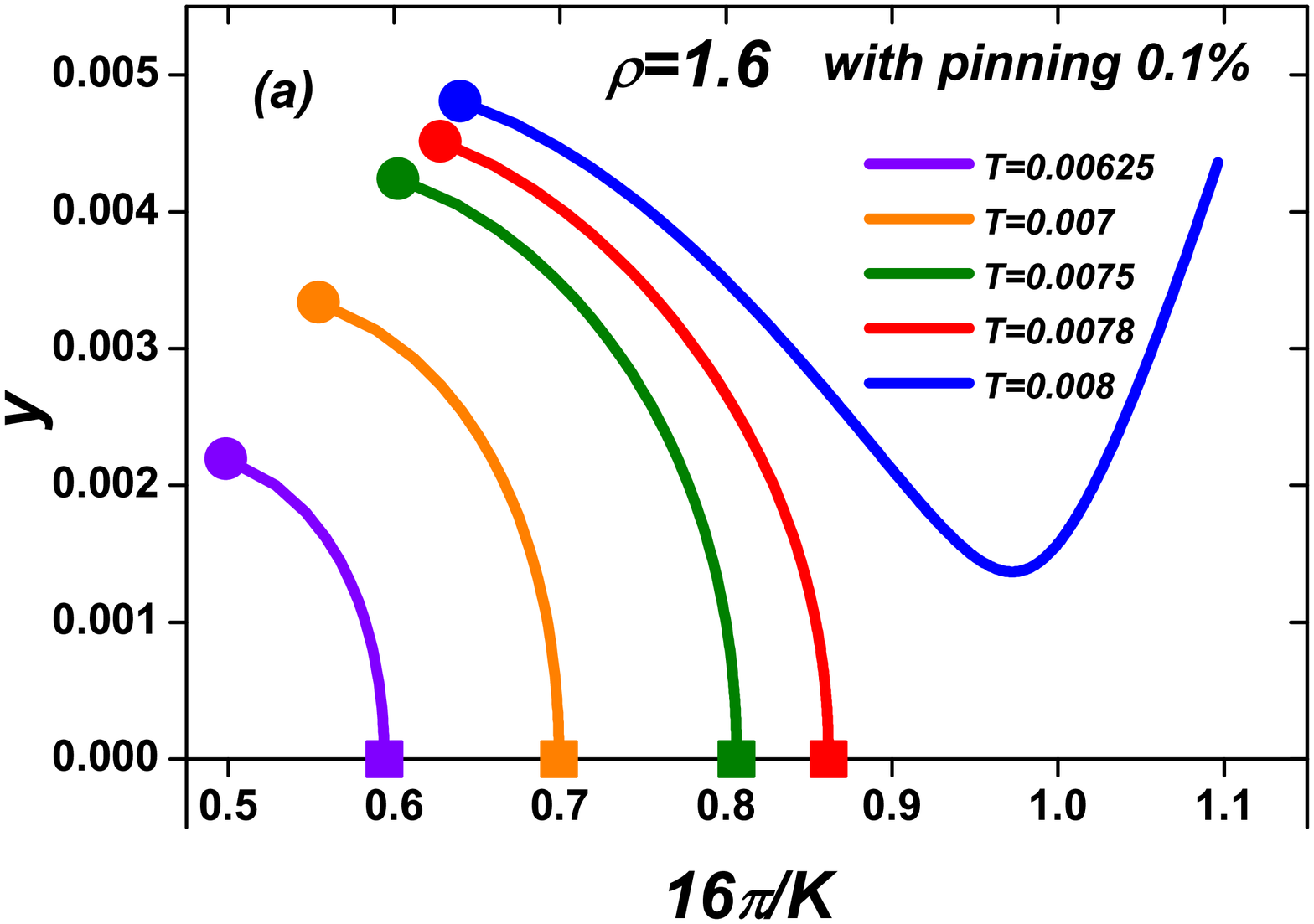}%

\includegraphics[width=8cm]{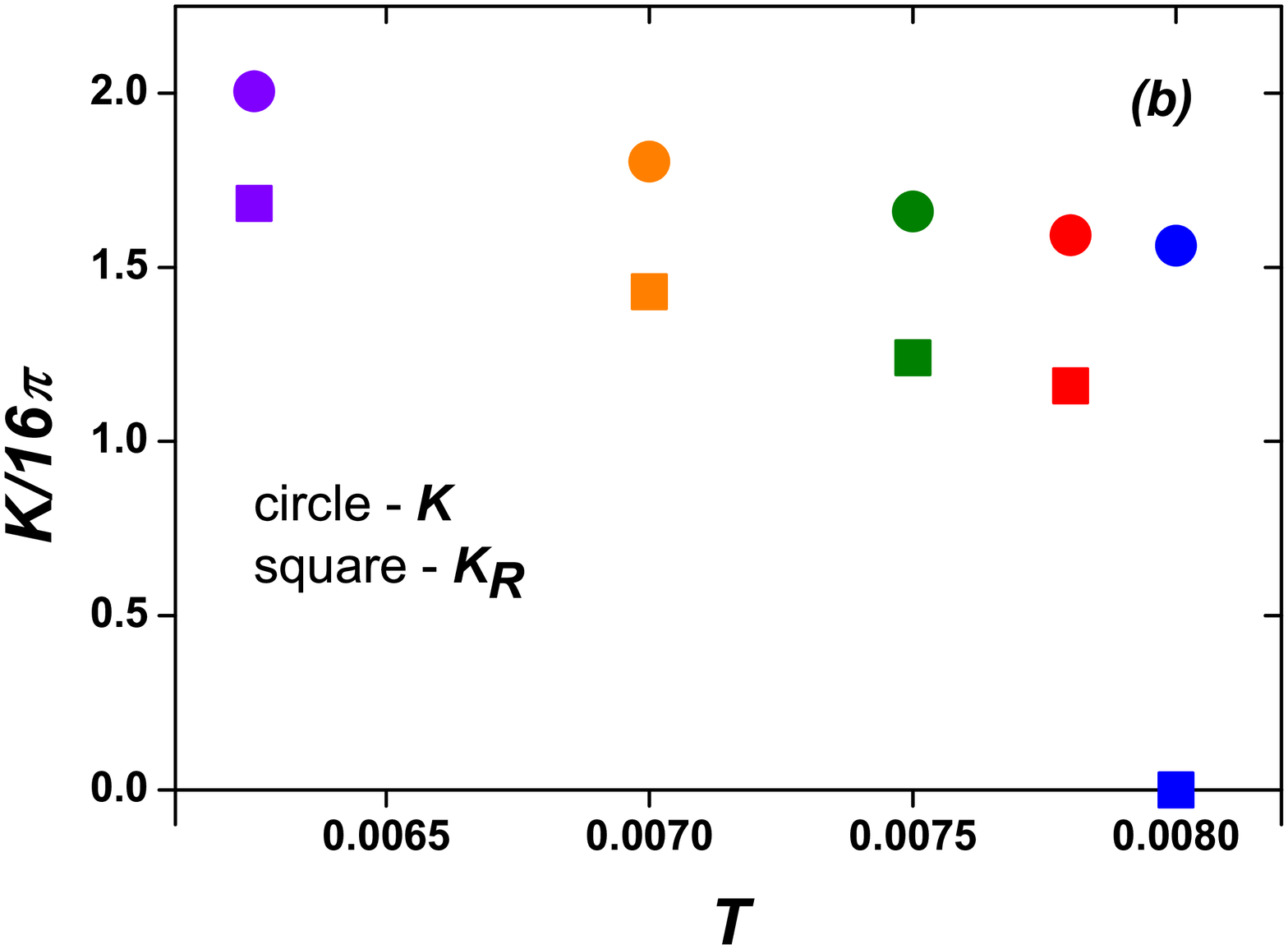}%

\includegraphics[width=8cm]{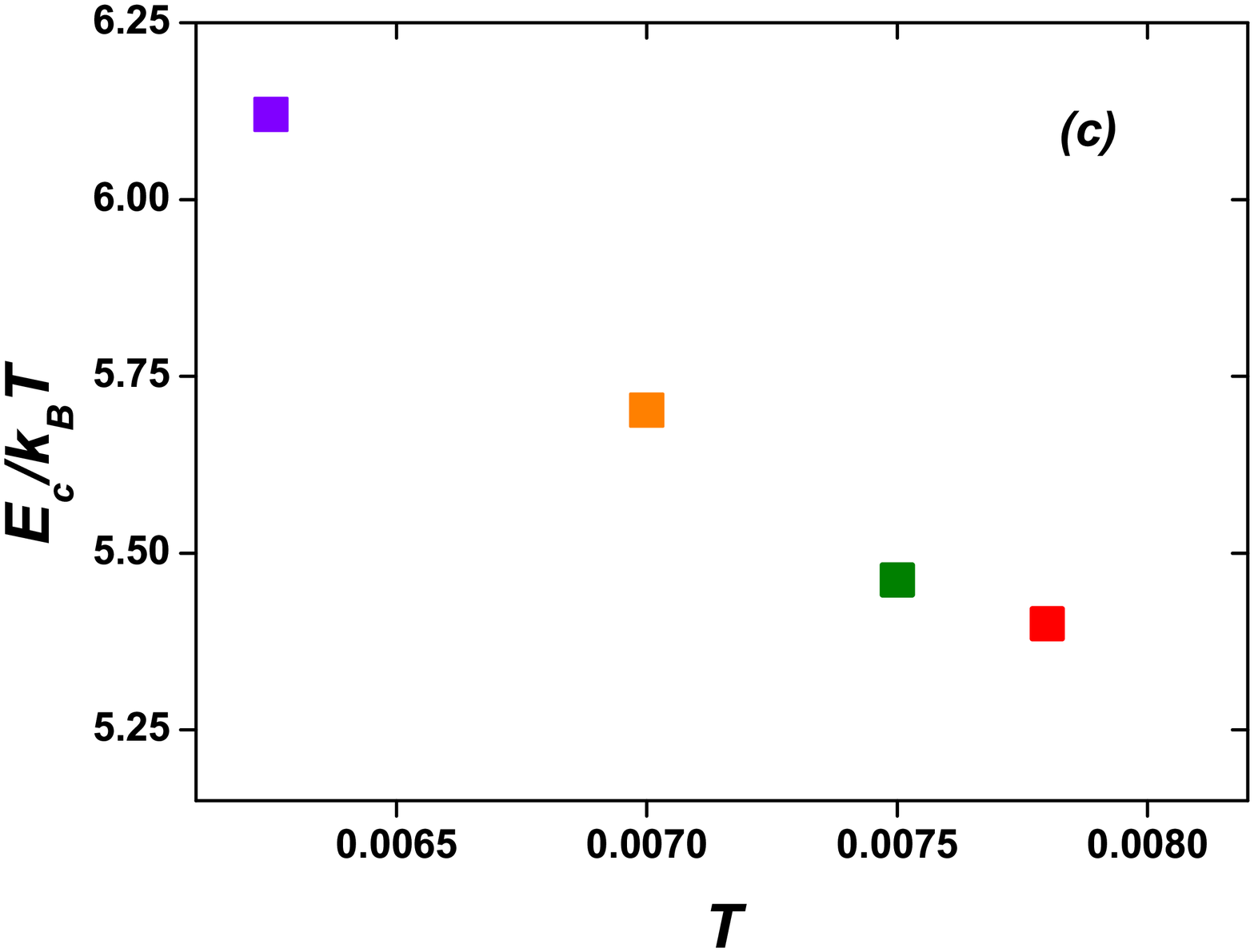}%
\caption{\label{muec-r16} (a) Trajectory $y-K$ as a result of
renormalization for different temperatures at density $\rho =1.6$
in the system with pinning $0.1$ \%. (b) The unrenormalized $K$
and renormalized $K_R$ Young moduli depending on temperature for
the same system. (c) Dimensionless energy of the dislocation core
$E_c/k_BT$ depending on temperature for the same system.}
\end{figure}

Fig.~\ref{muec-r16} (a) shows the trajectories in plane $y$-$K$
for different temperatures at density $\rho =1.6$. The crystal
loses stability when the curves leave for infinity generating an
unordered hexatic phase. From Fig.~\ref{muec-r16} (a) it is
evident that this happens at temperature between $T=0.0078$ and
$T=0.008$. At the same time according to the BKTHNY theory,
renormalized Young's modulus $K_R$ undergoes a sharp jump from
$16\pi$ to 0, which is caused by the loss of shear resistivity. As
shown in Fig.~\ref{muec-r16} (b) the unrenormalized $K$ and
renormalized $K_R$ Young moduli decrease with temperature growth
up to $T=0.0078$. Further temperature growth leads to a sharp fall
of $K_R$ to 0, i.e., the system transforms into the hexatic phase,
which is in good agreement with the result from $G_t$. It is
possible to conclude that both criteria are sensitive to formation
of dislocation pairs. The dimensionless energy of the dislocation
core shown in Fig.~\ref{muec-r16} (c) at the crystal-hexatic
transition point has value $E_c/k_BT = 5.4$, which is higher than
$2.84k_BT$ in the case of first-order transition \cite{chui}.
Hence, crystal-hexatic transition is due to dissociation of
dislocation pairs and is a continuous transition of the BKT type.
So, from the collection of all criteria it is possible to say with
confidence that in the system with pinning on tricritical isochore
$\rho=1.6$ the triangular crystal melts into the hexatic phase in
accordance with the BKT theory. Random pinning significantly
expanded the hexatic phase area at the expense of destruction of
the crystal but in no way did it affect the magnitude of
tricritical temperature.

\end{document}